\documentclass[10pt,conference]{IEEEtran}
\usepackage{cite}
\usepackage{amsmath,amssymb,amsfonts}
\usepackage{graphicx}
\usepackage{textcomp}
\usepackage{xcolor}
\usepackage[hyphens]{url}

\usepackage{algorithm}
\usepackage{algpseudocode}
\usepackage{pifont}
\usepackage[hidelinks]{hyperref}

\title{Rethinking ANNS on Commodity PIM: PIM-Native Vector Search with Cooperative Filtering}

\newcommand{\hpcaauthors}{%
  Dawei Guo,
  Yizou Chen,
  Fan Yang,
  Hui Li,
  and Ming-Chang Yang%
}

\newcommand{\hpcaaffiliation}{%
  The Chinese University of Hong Kong%
}

\newcommand{\hpcaemail}{%
  \{dwguo25, chenyz, fyang24, hli, mcyang\}@cse.cuhk.edu.hk%
}

\author{%
  \IEEEauthorblockN{\hpcaauthors}
  \IEEEauthorblockA{%
    \hpcaaffiliation\\
    \hpcaemail
  }
}

\begin{document}

\maketitle

\pagestyle{plain}
\thispagestyle{plain}

\newcommand{\sys}{PIVOT}
\setlength{\floatsep}{6pt plus 1pt minus 2pt}
\setlength{\textfloatsep}{6pt plus 1pt minus 2pt}
\setlength{\intextsep}{7pt plus 1pt minus 2pt}
\setlength{\abovecaptionskip}{2pt}
\newcommand{\tabitem}{%
    \leavevmode
    \textbullet\hspace{0.35em}%
}


\begin{abstract}
Approximate nearest neighbor search (ANNS) is a key primitive in modern data-intensive systems, and its memory-bound vector-scanning stage makes processing-in-memory (PIM) a promising acceleration platform. Existing PIM-based ANNS systems largely transplant CPU-oriented distance evaluation and top-$k$ selection onto PIM, introducing arithmetic- and state-heavy operations that poorly match lightweight, isolated PIM cores.
To address these mismatches, we present \sys{}, a \underline{PI}M-native
\underline{V}ector search system with c\underline{O}operative
fil\underline{T}ering, which thoroughly rethinks how ANNS should be executed on commodity PIM. First, \sys{} recasts PIM-side distance evaluation around binary quantization to exploit PIM-native bitwise operations, enabling LUT-free evaluation with low arithmetic cost and a small local-memory footprint. Second, \sys{} redesigns PIM-side candidate selection around threshold-guided cooperative filtering, using a survivor-mass model to derive a conservative threshold for CPU-side cluster filtering and PIM-side lower-bound filtering, thereby reducing output traffic and eliminating per-core top-$k$ state. \sys{} organizes these stages into a cooperative workflow that keeps query-global coordination and reranking on the CPU while assigning memory-intensive, data-parallel filtering to PIM.
To efficiently realize this workflow and sustain high utilization across devices, \sys{} further generates configuration-specialized PIM kernels and pipelines CPU-side preparation and reranking with PIM-side filtering across query batches. Evaluations on real UPMEM hardware show that \sys{} filters out 99.3--99.7\% of probed vectors and achieves up to $18.8\times$ and $10.2\times$ the QPS of CPU and prior PIM baselines, respectively. \sys{} also achieves higher recall than existing PIM systems and remains effective on high-dimensional embeddings.
\end{abstract}

\section{Introduction}
\label{sec:introduction}

Approximate nearest neighbor search (ANNS) has become a key primitive in modern data-intensive systems, including retrieval-augmented generation (RAG)~\cite{lewis2020rag,guu2020realm,karpukhin2020dpr}, vector databases~\cite{wang2021milvus,guo2022manu,zhang2023vbase}, recommendation systems~\cite{covington2016youtube,yi2019sampling}, and large-scale semantic retrieval~\cite{xiong2021ance,reimers2019sbert,santhanam2022colbertv2}. Existing ANNS methods are commonly grouped into cluster-based, graph-based, and hashing-based families~\cite{andoni2018annsurvey,wang2021graphsurvey,cai2021hashing}. Among them, cluster-based methods are widely used in billion-scale and production systems, most commonly in the form of inverted-file (IVF) indexes, due to their compact index structures, controllable search cost, and sequential, highly parallel scan patterns~\cite{jegou2011product,johnson2019billion,wang2021milvus}. At large scale, however, cluster-based ANNS spends substantial time scanning IVF lists, placing heavy pressure on memory bandwidth and exposing the von Neumann bottleneck~\cite{wulf1995memory,chen2024upanns,chen2025drimann,wu2025pimann,liu2025upvss}. Processing-in-memory (PIM) therefore offers a promising way to accelerate large-scale cluster-based ANNS by executing this scan close to data~\cite{ghose2018enabling,mutlu2019processing,gomezluna2022benchmarking}.

Commodity PIM systems place many lightweight processing units (PUs) close to memory, reducing data movement and providing high aggregate memory bandwidth~\cite{devaux2019truepim,lee2021hbmpim,kwon2022gddr6aim,gomezluna2022benchmarking}. Their architecture, however, differs substantially from that of CPUs. Because compute logic must coexist with dense memory arrays under stringent area, power, thermal, and timing constraints, commodity PIM typically integrates simple memory-side cores with modest arithmetic capability and limited local working memory for runtime state and buffering. PUs also operate independently, with no shared memory or direct inter-PU communication, making global synchronization and distributed state management costly~\cite{devaux2019truepim,gomezluna2022benchmarking}. These characteristics limit the effectiveness of directly transplanting CPU-oriented workflows. Fully exploiting PIM therefore requires a PIM-native execution path that keeps query-global control on the CPU and maps arithmetic-light, state-light, and independently executable work to the PUs.

\begin{table}[t]
\centering
\caption{\textbf{\sys{} at a glance.}}
\vspace{-3pt}
\label{tab:overview}
\small
\setlength{\tabcolsep}{3pt}
\begin{tabular}{
    @{}p{0.17\columnwidth}
    p{0.32\columnwidth}
    p{0.45\columnwidth}@{}
}
    \hline
    &
    \centering
    \textbf{Existing PIM ANNS}
    &
    \centering
    \textbf{\sys{}}
    \tabularnewline
    \hline

    \textbf{Evaluation Kernel Mismatch}
    &
    \textbf{--}\ FP/MUL-heavy
    \par
    \textbf{--}\ LUT-dependent
    &
    \textbf{+}\ Bitwise-dominated
    \par
    \textbf{+}\ LUT-free
    \\

    \hline

    \textbf{Search Paradigm Mismatch}
    &
    \textbf{--}\ Distributed top-$k$
    \par
    \textbf{--}\ Replicated state
    \par
    \textbf{--}\ Synchronization
    &
    \textbf{+}\ Cooperative filtering
    \par
    \textbf{+}\ Ranking-state-free
    \par
    \textbf{+}\ Independent per-PU filtering
    \\

    \hline
\end{tabular}
\end{table}

Existing PIM-based ANNS systems have demonstrated the potential of memory-side acceleration for large-scale vector search~\cite{chen2024upanns,chen2025drimann,wu2025pimann,liu2025upvss}. These systems adapt conventional IVF-based cluster-search pipelines to PIM through PIM-specific designs for data placement, query scheduling, and resource management. However, their vector-scanning and candidate-selection workflow still retains two CPU-oriented design choices that are poorly matched to commodity PIM architectures, as summarized in Table~\ref{tab:overview} and examined in Section~\ref{sec:investigation}. First, distance evaluation often relies on floating-point or multiplication-intensive computation and, in PQ-based designs, query-dependent lookup tables, consuming substantial processing cycles and local-memory capacity. Second, candidate selection retains a top-$k$ ranking paradigm designed for shared-memory CPUs, where ranking state can be efficiently managed in shared memory. On PIM, however, distributing a query across isolated PUs decomposes the global ranking into replicated local top-$k$ buffers with synchronized updates, causing ranking state and partial-result traffic to grow with the number of participating PUs.

These observations lead us to rethink ANNS execution on commodity PIM. We argue that a PIM-native ANNS pipeline should follow two principles. First, the PIM-side distance-evaluation kernel should favor arithmetic-light, PIM-native operations and compact runtime state. Second, candidate selection should replace distributed top-$k$ ranking with a query-specific threshold that enables stateless, independent filtering across PUs. Realizing these principles requires careful CPU--PIM work partitioning according to the strengths of each side.

Based on these principles, we propose \sys{}, a \underline{PI}M-native \underline{V}ector search system with c\underline{O}operative fil\underline{T}ering. 
For PIM-native distance evaluation, \sys{} builds on RaBitQ-style binary quantization and redesigns the corresponding distance-evaluation kernel with centered query quantization and reduced-precision FastFloat reconstruction, yielding a LUT-free calculation dominated by PIM-native bitwise and integer operations~\cite{gao2024rabitq,gao2025extrabitq,faisscentered}.
To eliminate distributed top-$k$ ranking, \sys{} replaces ranking with threshold-guided cooperative filtering. Specifically, \sys{} introduces a query-aware threshold estimator that combines per-cluster residual statistics with query-to-centroid distances to derive a conservative query-specific radius $\tau$ and estimate the expected survivor contribution of each probed cluster. Using these estimates, the CPU first suppresses low-yield clusters before dispatch. The PUs then stream PIM-resident compact codes, independently reject candidates whose squared-distance lower bounds exceed $\tau^2$, and return only survivor identifiers for CPU reranking. \sys{} further pipelines consecutive query batches, overlapping CPU-side routing, threshold estimation, and reranking with PIM-side filtering. Together, these mechanisms address the two architectural mismatches while keeping the PIM-side hot path arithmetic-light, state-light, and streaming-oriented.

This paper makes the following contributions:
\begin{itemize}
    \item We identify two architectural mismatches in existing PIM-based cluster search: arithmetic-heavy, LUT-dependent distance evaluation, and distributed top-$k$ ranking with synchronization, replicated state, and communication overhead (Section~\ref{sec:investigation}).
    
    \item We propose \sys{}, a PIM-native ANNS system with cooperative filtering. It combines a PIM-native lower-bound evaluation kernel with a query-aware threshold estimator to enable cooperative CPU-side cluster filtering and independent PIM-side vector filtering (Section~\ref{sec:design}).
    
    \item We implement \sys{} on UPMEM with configuration-specialized PIM kernels and pipelined CPU--PIM execution (Section~\ref{sec:implementation}).

    \item We evaluate \sys{} on billion-scale datasets. Unlike prior PIM systems whose PQ and local top-$k$ designs limit scalability toward practical high-recall operation, \sys{} consistently delivers a stronger throughput--recall tradeoff, with its throughput advantage widening as recall increases. It achieves up to $18.8\times$ and $10.2\times$ the QPS of the CPU and PIM baselines, respectively, making commodity PIM more practical for real-world ANNS workloads (Section~\ref{sec:evaluation}).
\end{itemize}

\section{Background}
\label{sec:background}

\subsection{Commodity PIM Architecture}
\label{sec:bg-pim}

\begin{figure}[t]
\centering
\includegraphics[width=0.8\columnwidth]{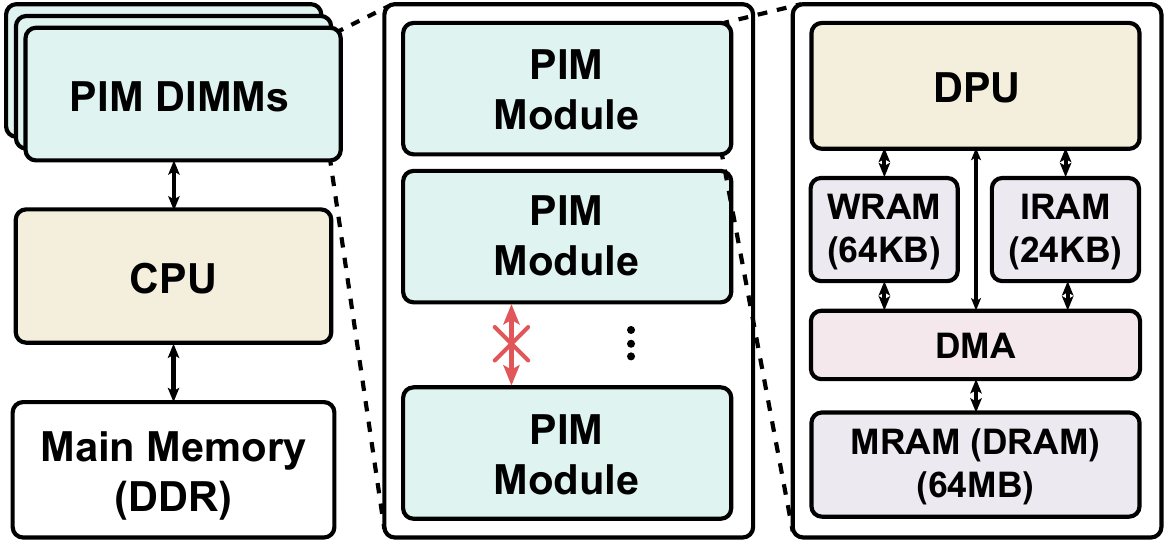}
\caption{\textbf{A representative commodity PIM architecture: UPMEM PIM.}}
\label{fig:pim_arch}
\vspace{-6pt}
\end{figure}

Commodity PIM systems integrate many lightweight processing units (PUs) close to DRAM banks to reduce off-chip data movement and provide high aggregate memory bandwidth~\cite{devaux2019truepim,lee2021hbmpim,kwon2022gddr6aim}. UPMEM PIM~\cite{devaux2019truepim}, illustrated in Figure~\ref{fig:pim_arch}, is the first publicly available programmable commodity DRAM-PIM system and the hardware substrate used by most prior real-hardware PIM-based ANNS systems~\cite{chen2024upanns,chen2025drimann,wu2025pimann,liu2025upvss}. We therefore use UPMEM as a representative evaluation platform, while our design principles target broader commodity-PIM architectural tradeoffs and extend beyond this specific platform.

A UPMEM-enabled system consists of a host CPU and multiple PIM DIMMs connected through standard memory channels. Each PIM DIMM integrates many DRAM Processing Units (DPUs), each operating on a private 64~MB DRAM bank referred to as MRAM. A DPU also contains a lightweight in-order processor, 64~KB of software-managed scratchpad memory (WRAM) for local state and buffering, and 24~KB of instruction memory (IRAM). Multiple tasklets share the DPU pipeline and WRAM, with their execution interleaved to improve pipeline utilization and hide memory-access latency~\cite{devaux2019truepim,gomezluna2022benchmarking,hyun2024pathfinding}.

Because compute logic must coexist with dense memory arrays under tight area, power, thermal, and timing budgets, commodity PIM trades per-core complexity, fast local storage, and communication flexibility for large-scale memory-side parallelism~\cite{mutlu2022primer,lee2021hbmpim,kwon2022gddr6aim}. These tradeoffs directly shape the PIM programming model:

\textbf{Modest arithmetic capability:}
Complex arithmetic units consume additional area and power, so memory-side cores favor relatively simple instruction sets. On UPMEM, integer addition, bitwise operations, and shifts are efficiently supported, whereas floating-point operations are emulated in software and full-width multiplication and division require longer instruction sequences~\cite{gomezluna2022benchmarking,upmem2025codingtips}. Kernels dominated by floating-point or multiplication-intensive computation can therefore quickly exhaust DPU arithmetic throughput~\cite{rhyner2024pimopt}.

\textbf{Limited local runtime memory:}
SRAM is substantially less dense than DRAM and becomes costly when replicated across many memory-side cores. Commodity PIM therefore provides only a small amount of fast local memory, such as WRAM on UPMEM. Data processed by a DPU must be explicitly transferred between MRAM and WRAM through DMA operations~\cite{devaux2019truepim,hyun2024pathfinding}. WRAM is shared by all tasklets and must accommodate tasklet stacks, temporary buffers, and runtime state. Large lookup tables or replicated per-tasklet buffers can therefore exhaust WRAM, reducing tasklet concurrency and limiting supported workload configurations~\cite{gomezluna2022benchmarking,diab2023aim,chen2024upanns}.

\textbf{Restricted communication and coordination:}
Providing coherent shared memory or direct communication among thousands of memory-side cores would incur substantial hardware overhead~\cite{noh2024pidcomm}. Consequently, different DPUs do not share memory or communicate directly. Global reductions, shared-state updates, and inter-DPU data exchange must be mediated by the host CPU through explicit transfers~\cite{kang2022pimtree}. Within a DPU, UPMEM supports synchronization among tasklets, although frequent synchronization can reduce tasklet-level parallelism~\cite{gomezluna2022benchmarking,upmem2025codingtips}.

These characteristics create a programming model substantially different from conventional CPU execution. Efficient commodity-PIM execution therefore requires rethinking application workflows and the CPU--PIM boundary, keeping global control and coordination on the CPU while structuring PIM-side execution around streaming, arithmetic-light, and low-state kernels~\cite{kang2022pimtree}.

\subsection{Cluster-Based ANNS}
\label{sec:bg-anns}

Given a query vector $q$, top-$k$ ANNS seeks to return the $k$ database vectors closest to $q$. Cluster-based methods commonly organize the search using an IVF index~\cite{jegou2011product,johnson2019billion}. During indexing, database vectors are partitioned into $nlist$ clusters, typically using $k$-means, and each cluster is represented by a centroid and an inverted list. The vectors are typically encoded into quantized representations to reduce index size and distance-evaluation cost. At query time, search typically proceeds in three stages. First, \textbf{coarse routing} computes query-to-centroid distances and selects the nearest $nprobe$ lists. Second, \textbf{vector scanning and candidate selection} evaluates distances using the quantized representations in these lists and retains the closest candidates in a bounded pool. Third, because quantization introduces approximation error, \textbf{reranking} reevaluates this pool using a more accurate representation, such as a less compressed quantized representation or the original full-precision vectors, before returning the final results.

\textbf{Vector quantization and evaluation kernels:}
Direct L2 evaluation over full-precision vectors performs floating-point subtraction, multiplication, and accumulation across every dimension. Scalar quantization reduces storage and computation costs by mapping each coordinate to a low-precision value, such as INT8. To achieve higher compression and faster evaluation, product quantization (PQ) divides each vector into $M$ subvectors and encodes each with its nearest codeword~\cite{jegou2011product}. For each query, PQ constructs query-dependent lookup tables and evaluates each vector through $M$ lookups and accumulations. Binary quantization methods such as RaBitQ further compress vectors into compact bit strings and estimate distances mainly through bitwise operations and population counts, together with limited auxiliary metadata~\cite{gao2024rabitq,gao2025extrabitq}. RaBitQ also provides a probabilistic error bound at a specified confidence level, enabling a conservative lower bound on the true squared Euclidean distance to be derived.

\textbf{Top-$k$ candidate selection:}
Conventional CPU-based IVF search follows a score-then-rank workflow. As vector distances are evaluated, the system maintains a bounded max-heap containing the closest candidates found so far. The heap is updated whenever a closer candidate is discovered. Bound-based methods such as RaBitQ can use the largest retained distance at the heap root as a dynamic pruning threshold. If a vector's distance lower bound exceeds this threshold, it cannot enter the candidate pool, allowing further evaluation or reranking to be skipped~\cite{gao2024rabitq,gao2025extrabitq}. When reranking is used, the system retains more than $k$ candidates, reevaluates them using a higher-fidelity representation, and selects the final top-$k$ results according to the refined distances.

\subsection{Existing ANNS Systems on Commodity PIM}
\label{sec:bg-pim-anns}

The vector-scanning stage of cluster-based ANNS generates substantial memory traffic and exposes abundant parallelism across inverted lists and their vectors, making it a natural target for commodity PIM. Existing PIM-based ANNS systems follow this approach by accelerating the scan stage of IVF-based pipelines on memory-side processors~\cite{chen2024upanns,chen2025drimann,wu2025pimann,liu2025upvss}, while retaining query-global operations such as coarse routing and final result aggregation on the CPU. These systems differ primarily in the representation and kernel used for vector evaluation and in how top-$k$ candidate selection is organized.

UPVSS performs floating-point L2 distance evaluation on the PIM side and returns all evaluated vector distances and identifiers to the CPU for global top-$k$ selection~\cite{liu2025upvss}. Under this design, the DPUs serve as parallel distance-evaluation engines, while candidate selection remains entirely on the host. UpANNS and DRIM-ANN instead use PQ-based vector evaluation and perform partial top-$k$ candidate selection within the DPUs~\cite{chen2024upanns,chen2025drimann}. UpANNS maintains a private candidate heap for each tasklet and aggregates the tasklet-local heaps into a DPU-level result. DRIM-ANN maintains a shared DPU-level top-$k$ priority queue that is cooperatively updated by multiple tasklets. In both systems, the partial candidate sets produced by different DPUs are subsequently merged by the CPU.

Existing PIM-based ANNS systems retain the conventional score-then-rank workflow. They either return all evaluated vector distances for CPU-side top-$k$ selection or maintain partial ranking state on the PIM side. Section~\ref{sec:investigation} examines the resulting evaluation--kernel and search--paradigm mismatches under the arithmetic, WRAM, communication, and coordination constraints of commodity PIM.

\section{Investigation}
\label{sec:investigation}

Existing PIM-based ANNS systems demonstrate that cluster-based ANNS can benefit from memory-side execution~\cite{chen2024upanns,chen2025drimann,liu2025upvss}. Their vector-scanning and candidate-selection paths, however, preserve the CPU-oriented score-then-rank workflow, exposing two mismatches on commodity PIM.

\subsection{Evaluation--Kernel Mismatch: CPU-Friendly Kernels Are Not PIM-Friendly}
\label{sec:investigation-representation}

The first mismatch arises from the quantization methods and corresponding distance-evaluation kernels used during vector scanning. CPU-oriented IVF systems typically select quantization methods based on storage efficiency and evaluation speed on cache-rich, arithmetic-capable processors. On commodity PIM, kernel efficiency also depends on whether its arithmetic operations match lightweight memory-side cores and whether its query-dependent state fits within limited local memory.

\begin{figure}[t]
\centering
\includegraphics[width=\linewidth]{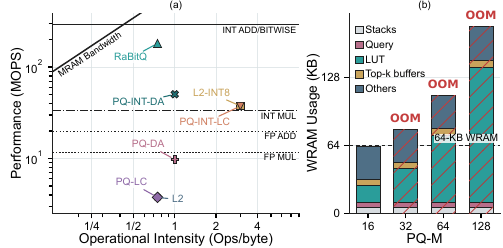}
\vspace{-12pt}
\caption{\textbf{Evaluation--kernel mismatch on commodity PIM.}
(a) Empirical roofline of representative distance-evaluation kernels on PIM DPUs.
\textbf{L2}: full-precision L2. \textbf{L2-INT8}: INT8 L2.
\textbf{PQ-LC/DA}: PQ lookup-table construction and distance accumulation.
\textbf{PQ-INT-LC/DA}: integer-quantized PQ variants.
\textbf{RaBitQ}: RaBitQ-style XOR--population-count kernel.
(b) WRAM usage of PQ execution as the number of subquantizers $M$ increases.
The dashed line marks the 64\,KB WRAM capacity.}
\label{fig:investigation-evaluation}
\end{figure}

Figure~\ref{fig:investigation-evaluation}(a) presents an empirical roofline of representative distance-evaluation kernels on UPMEM. The $x$-axis reports effective operations per MRAM byte, while the $y$-axis reports effective throughput in millions of operations per second (MOPS). The diagonal MRAM-bandwidth roof and horizontal throughput ceilings for different DPU operation classes are measured on real UPMEM hardware.

Conventional distance-evaluation kernels fall well below the PIM performance roof because their arithmetic requirements poorly match lightweight memory-side cores. Full-precision L2 performs poorly because floating-point operations are emulated in software. INT8 L2 removes floating point but remains multiplication-intensive. PQ reduces the bytes scanned per vector, but lookup-table construction (PQ-LC) resembles dimension-wise L2 evaluation, while floating-point distance accumulation (PQ-DA) remains expensive. Integer-quantized PQ improves throughput, but still requires multiplication-intensive LUT construction and introduces additional quantization error. In contrast, binary distance evaluation relies primarily on bitwise operations and population counts, moving execution closer to the performance roof.

The mismatch also appears in WRAM usage. A PQ lookup table with $M$ subquantizers and $2^{nbits}$ entries per subquantizer requires
$M \cdot 2^{nbits} \cdot s_{\mathrm{entry}}$ bytes, where $s_{\mathrm{entry}}$ is the entry size. With the common setting $nbits=8$ and 32-bit entries, the LUT alone occupies $1024M$ bytes. Figure~\ref{fig:investigation-evaluation}(b) shows that LUTs dominate the WRAM footprint and grow linearly with $M$. Total WRAM usage approaches UPMEM's 64\,KB capacity at $M=16$ and exceeds it from $M=32$. Because larger $M$ is often required for high-dimensional vectors and higher accuracy, PQ creates a direct tension between search accuracy and WRAM capacity.

These results show that the vector representation and its complete distance-evaluation path must be designed jointly for commodity PIM. Binary quantization avoids query-dependent LUTs and enables a bitwise scan kernel that better matches DPU arithmetic and local-memory constraints.

\subsection{Search--Paradigm Mismatch: Distributed Top-\texorpdfstring{$k$}{k} Selection Is Not PIM-Friendly}
\label{sec:investigation-paradigm}

The second mismatch stems from the stateful nature of top-$k$ candidate selection. On CPUs, worker threads share memory and can efficiently manage ranking state for the current top-$k$ candidates. On commodity PIM, exploiting thousands of isolated DPUs requires distributing each query's scan across DPUs, with multiple tasklets processing vectors within each DPU. Because DPUs neither share memory nor communicate directly, PIM-side top-$k$ selection must be decomposed into local rankings across DPUs or tasklets, followed by host-side aggregation. We refer to this paradigm as \emph{PIM-side distributed top-$k$}.

\begin{figure}[t]
    \centering
    \includegraphics[width=\linewidth]{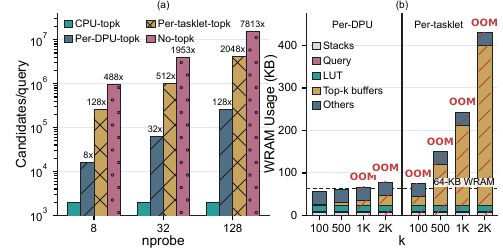}
    \vspace{-12pt}
    \caption{\textbf{Search--paradigm mismatch on commodity PIM.}
    (a) Returned candidates per query under different selection methods. Labels show amplification relative to CPU-global top-$k$.
    (b) WRAM usage of per-DPU and per-tasklet top-$k$ as $k$ increases. The dashed line marks the 64\,KB WRAM capacity.}
    \label{fig:investigation-topk}
\end{figure}

Per-DPU top-$k$, as used by DRIM-ANN~\cite{chen2025drimann}, maintains one shared heap in each participating DPU. Across $P$ DPUs, these heaps collectively retain up to $P \cdot k$ candidates for CPU aggregation. As shown in Figure~\ref{fig:investigation-topk}(a), this produces substantially more output than CPU-global top-$k$. Multiple tasklets within each DPU must also synchronize when updating the shared heap, introducing serialized accesses into the scan hot loop.

Per-tasklet top-$k$, as used by UpANNS~\cite{chen2024upanns}, assigns each tasklet a private heap to avoid shared-heap synchronization. With $T$ tasklets per DPU, the retained capacity grows to $P \cdot T \cdot k$, amplifying candidate output while increasing WRAM usage, PIM--CPU transfer, and CPU-side merge work.

Figure~\ref{fig:investigation-topk}(b) shows that both organizations quickly exhaust local memory as $k$ increases. The per-DPU design exceeds UPMEM's 64\,KB WRAM capacity at $k=1000$, while the per-tasklet design exceeds it at $k=100$ because ranking state is replicated across tasklets. This leaves insufficient space for lookup tables, stacks, DMA buffers, and other runtime data, limiting the supported $k$ and workload configurations.

An alternative is to omit PIM-side top-$k$ selection and return the distance of every scanned vector to the CPU, as in UPVSS~\cite{liu2025upvss}. This eliminates PIM-side ranking state and synchronization but transfers excessive data. Figure~\ref{fig:investigation-topk}(a) shows that the returned candidate count grows rapidly with $nprobe$, reaching up to $7813\times$ that of CPU-global top-$k$. The resulting PIM--CPU transfer and CPU-side selection overhead can offset the benefits of memory-side scanning.

Distributed top-$k$ also loses the globally updated pruning threshold available in shared-memory CPU execution. In lower-bound methods such as RaBitQ, the largest distance in the current heap serves as a dynamic cutoff, allowing vectors whose lower bounds exceed it to be discarded early. Under PIM-side distributed top-$k$, each DPU observes only its local ranking state, while maintaining a global cutoff would require repeated host-mediated communication during scanning. When all distances are returned to the CPU, the global cutoff becomes available only after scanning and transfer are complete. Neither organization therefore provides a timely global cutoff for effective lower-bound pruning on PIM.

PIM-side distributed top-$k$ incurs synchronization, replicated ranking state, and candidate-output amplification, while host-side global top-$k$ incurs excessive communication and CPU-side selection overhead. These limitations motivate independent threshold-guided filtering on PIM, with global ranking performed only over surviving candidates on the CPU.

\section{\sys{} Design}
\label{sec:design}

\subsection{Design Principles}
\label{sec:design-principles}

Section~\ref{sec:investigation} identifies an evaluation--kernel mismatch and a search--paradigm mismatch when CPU-oriented ANNS workflows are mapped onto commodity PIM. \sys{} follows three design principles to address these mismatches.

\textbf{PIM-native distance-evaluation kernel:}
The evaluation--kernel mismatch shows that PIM-side distance evaluation should be arithmetic-light and state-light. \sys{} recasts distance evaluation around RaBitQ-style binary quantization, enabling LUT-free computation dominated by XOR and population-count operations~\cite{gao2024rabitq,gao2025extrabitq}. It further uses centered query quantization to reduce binary-side arithmetic and introduces reduced-precision FastFloat to accelerate auxiliary floating-point reconstruction. Together, these choices yield a PIM-native lower-bound evaluation kernel whose per-vector path is dominated by PIM-native operations (Section~\ref{sec:design-kernel}).

\textbf{Cooperative filtering instead of distributed top-$k$:}
The search--paradigm mismatch shows that PIM-side distributed top-$k$ incurs synchronization, replicated ranking state, and candidate-output amplification. \sys{} instead derives a query-specific radius $\tau(q)$ that enables independent, ranking-state-free filtering across DPUs. Through threshold-guided cooperative filtering, the CPU filters low-yield clusters before dispatch, while each DPU independently rejects vectors whose squared-distance lower bounds exceed $\tau^2(q)$ and returns only survivor identifiers to the CPU for reranking (Section~\ref{sec:design-filtering}).

\textbf{CPU--PIM work partitioning and pipelined execution:}
\sys{} partitions work according to each side's strengths. Query-global control, threshold estimation, cluster filtering, scheduling, and final reranking remain on the CPU, while the DPUs perform bulk vector scanning and lower-bound filtering. Consecutive query batches are pipelined so that CPU-side preparation and reranking overlap with PIM-side filtering.

\subsection{End-to-End Workflow}
\label{sec:design-workflow}

\begin{figure*}[t]
    \centering
    \includegraphics[width=\textwidth]{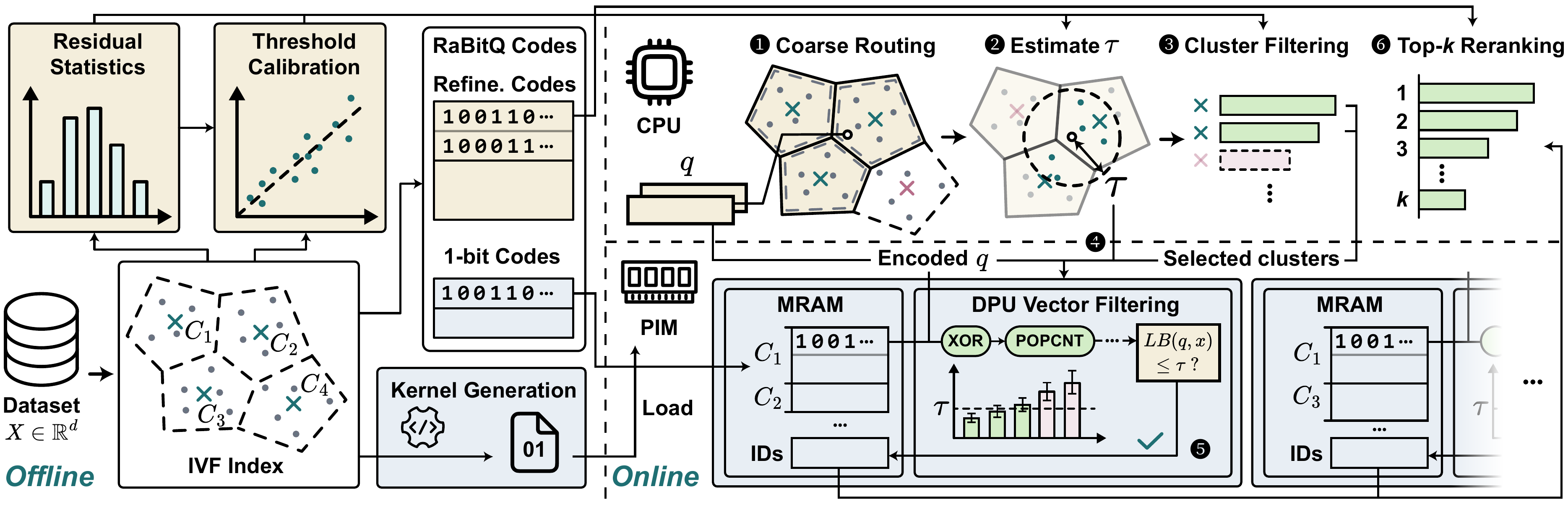}
    \vspace{-12pt}
    \caption{\textbf{\sys{} end-to-end workflow.}}
    \label{fig:filterpim-workflow}
    \vspace{-12pt}
\end{figure*}

Figure~\ref{fig:filterpim-workflow} presents the end-to-end workflow of \sys{}.

\textbf{Offline indexing phase:}
\sys{} constructs an IVF index with multi-bit RaBitQ codes following the original indexing procedure~\cite{gao2024rabitq,gao2025extrabitq}. Each vector contains a 1-bit base code for PIM-side lower-bound filtering and refinement bits for CPU reranking. For each cluster, \sys{} builds a residual-norm histogram for survivor-mass estimation and calibrates the threshold estimator using a small held-out query set. Sampled database vectors can serve as surrogate calibration queries when query traces are unavailable. The 1-bit codes and auxiliary factors are placed in PIM, while the complete multi-bit codes remain in host memory. Configuration-specialized DPU kernels are generated according to the dataset and index configuration and loaded before online execution. Section~\ref{sec:implementation} further describes the data layout and kernel specialization.

\textbf{Online search phase:}
Queries are processed in batches. \ding{182} The CPU performs IVF coarse routing and selects $nprobe$ clusters. \ding{183} The calibrated survivor-mass estimator derives a query-specific radius $\tau(q)$ and estimates the contribution of each selected cluster. \ding{184} Clusters with negligible expected survivor mass are removed before dispatch. \ding{185} The CPU packages the query metadata and squared threshold $\tau^2(q)$ and dispatches the remaining work to the resident DPUs. \ding{186} Each DPU applies the PIM-native lower-bound calculation to its local 1-bit codes and filters vectors against $\tau^2(q)$. \ding{187} The DPUs return survivor identifiers, which the CPU reranks using the complete multi-bit codes before selecting the final top-$k$ results. Consecutive batches are pipelined to overlap CPU-side routing, threshold estimation, and reranking with PIM-side filtering.

\subsection{PIM-Native Distance-Evaluation Kernel}
\label{sec:design-kernel}

RaBitQ-style binary quantization represents each database vector with a 1-bit code and derives a probabilistic squared-distance lower bound in two steps~\cite{gao2024rabitq,gao2025extrabitq}. It first computes an integer similarity term from the binary database code and quantized query, and then combines this term with per-vector and query--cluster auxiliary factors to reconstruct a squared-distance estimate and its probabilistic error bound. The binary calculation is well suited to PIM because it is LUT-free and dominated by bitwise operations. Directly mapping the complete distance path to a DPU, however, still requires a per-code population count and software-emulated floating-point reconstruction.

To reduce binary-side arithmetic, \sys{} adopts the centered-query formulation implemented in FAISS~\cite{faisscentered}, which quantizes the query around zero. This formulation eliminates the separate population count over each database code and computes the integer similarity term using weighted XOR--population-count operations and integer accumulation. The resulting value $I(q,x)$ is then used to reconstruct the squared-distance lower bound:
\[
    \mathrm{LB}_2(q,x)
    =
    a_x
    +
    \rho_j^2
    -
    b_x I(q,x)s_{q,j}
    -
    \rho_j e_x,
\]
where $a_x$, $b_x$, and $e_x$ are per-vector factors, $\rho_j=\lVert q-c_j\rVert_2$, and $s_{q,j}$ is the centered-query scale. At the configured confidence level, $\mathrm{LB}_2(q,x)$ lower-bounds $\lVert q-x\rVert_2^2$.

Evaluating this expression with UPMEM's general-purpose software-emulated floating-point routines for every vector remains expensive~\cite{gomezluna2022benchmarking,upmem2025codingtips}. \sys{} therefore introduces a reduced-precision \emph{FastFloat} path specialized to lower-bound reconstruction. FastFloat represents each operand using its sign, exponent, and a compile-time-configured reduced-width significand. It performs exponent alignment, significand arithmetic, and result normalization using integer and bitwise operations. It fuses $b_x I(q,x)s_{q,j}$ into a single reduced-precision operation, avoiding intermediate conversion and normalization. The resulting kernel combines centered XOR--population-count calculation with specialized FastFloat reconstruction, removing general software floating point from the per-vector path. Section~\ref{sec:eval-techniques} evaluates the performance and accuracy impact of FastFloat.

\subsection{Threshold-Guided Cooperative Filtering}
\label{sec:design-filtering}

The PIM-native kernel produces per-vector lower bounds, but using them across isolated DPUs requires a query-specific threshold before scanning and a filtering workflow that avoids distributed ranking state. \sys{} solves these challenges through threshold-guided cooperative filtering. The following describes (1) how \sys{} generates the threshold and (2) how it uses the threshold for CPU--PIM cooperative filtering.

\textbf{(1) Query-specific threshold generation:}
\sys{} estimates the smallest radius expected to contain the final top-$k$ vectors and calibrates it against modeling error.

\emph{Per-cluster residual statistics:}
Let $C_j$ denote an IVF cluster with centroid $c_j$. For each vector $x\in C_j$, define its residual as $z_x=x-c_j$ and its residual norm as $s_x=\lVert z_x\rVert_2$. During indexing, \sys{} summarizes these norms using an $N_h$-bin histogram
\[
    H_j=
    \left\{
        \left(s_{j,b},w_{j,b}\right)
    \right\}_{b=1}^{N_h},
\]
where $s_{j,b}$ is the mean residual norm in bin $b$ and $w_{j,b}$ is the fraction of cluster vectors assigned to that bin. Because cluster populations and residual norms are already available from the IVF--RaBitQ index, the estimator adds only $N_h$ entries per cluster and no per-vector metadata.

\emph{Query-conditioned survivor mass:}
For query $q$ and probed cluster $C_j$, define $u_j=q-c_j$ and $\rho_j=\lVert u_j\rVert_2$. A vector with residual norm $s$ has squared query distance
\[
    \lVert x-q\rVert_2^2
    =
    s^2+\rho_j^2-2s\rho_j\cos\theta,
\]
where $\theta$ is the angle between $z_x$ and $u_j$. Thus, for fixed $s$ and $\rho_j$, whether the vector lies within radius $r$ depends on their angular alignment.

\sys{} approximates residual directions within each histogram bin as isotropic. Under this approximation, the high-dimensional angular term satisfies $\sqrt{D}\cos\theta\overset{\mathrm{approx.}}{\sim}\mathcal{N}(0,1)$~\cite{stam1982sphere}. Let $\overline{\Phi}(\cdot)$ denote the upper-tail probability of the standard normal distribution. The estimated fraction and expected number of vectors in $C_j$ within radius $r$ are
\[
\begin{aligned}
    \widehat{G}_j(q,r)
    &=
    \sum_{b=1}^{N_h}
    w_{j,b}\,
    \overline{\Phi}
    \left(
        \sqrt{D}\,
        \frac{s_{j,b}^{2}+\rho_j^{2}-r^{2}}
             {2s_{j,b}\rho_j}
    \right), \\
    M_j(q,r)
    &=
    |C_j|\widehat{G}_j(q,r).
\end{aligned}
\]

\emph{Raw threshold selection:}
Let $\mathcal{P}(q)$ denote the clusters selected by coarse routing. The expected total survivor mass within radius $r$ is
\[\widehat{\Lambda}(q,r)=\sum_{C_j\in\mathcal{P}(q)}M_j(q,r).\]
\sys{} selects the smallest radius expected to contain $k$ vectors:
\[
    \tau_0(q)
    =
    \inf
    \left\{
        r
        \mid
        \widehat{\Lambda}(q,r)\ge k
    \right\}.
\]
Because $\widehat{\Lambda}(q,r)$ increases monotonically with $r$, $\tau_0(q)$ can be found efficiently through binary search.

\emph{Conservative threshold calibration:}
An overly large radius retains excess vectors, increasing PIM output and CPU reranking work, while an insufficient radius may exclude vectors needed for the final top-$k$ result. The isotropic approximation is intentionally lightweight and can introduce modeling error, so \sys{} calibrates $\tau_0(q)$ using a small offline query set. For each calibration query $q_i$, let $R_k(q_i)$ denote the exact distance to the $k$th closest vector among the same probed IVF lists. \sys{} selects a multiplicative correction factor $\beta_\epsilon$ such that
\[
    \Pr_{q_i\in Q_{\mathrm{cal}}}
    \left[
        R_k(q_i)>
        \beta_\epsilon\tau_0(q_i)
    \right]
    \le
    \epsilon,
    \qquad
    \tau(q)=\beta_\epsilon\tau_0(q).
\]
A smaller $\epsilon$ produces a more conservative threshold and retains more survivors. A larger $\epsilon$ reduces PIM output while increasing the risk of excluding vectors needed for the final top-$k$ result. Section~\ref{sec:eval-sensitivity} evaluates this tradeoff.

\textbf{(2) CPU--PIM cooperative filtering:}
The calibrated threshold $\tau(q)$ enables independent per-vector filtering on the DPUs. The survivor-mass model additionally uses this threshold to identify low-yield clusters, allowing the CPU to suppress them before dispatch.

\emph{CPU cluster-level filtering:}
For each cluster $C_j$ selected by IVF coarse routing, the CPU evaluates $M_j(q,\tau(q))$, the expected number of vectors within the calibrated radius, and dispatches the cluster only if $M_j(q,\tau(q))\ge\mu_{\min}$. Because survivor mass is estimated, $\mu_{\min}$ is calibrated for each supported $nprobe$ range so that $\mathrm{Recall@}k(\mu_{\min})\ge\mathrm{Recall@}k(0)-\delta$, where $\delta$ is the maximum allowed absolute recall loss from cluster filtering. Section~\ref{sec:eval-sensitivity} evaluates the tradeoff controlled by $\delta$. Compared with directly reducing $nprobe$, which enforces a smaller fixed probe count and discards later-ranked clusters solely by query-to-centroid distance, \sys{} evaluates every probed cluster using query-conditioned survivor mass. It adaptively suppresses low-yield clusters while preserving those likely to contribute useful survivors.

\emph{PIM vector-level filtering:}
The CPU dispatches each admitted list to its resident DPU together with the encoded query, squared threshold $\tau^2(q)$, and the factors required by the lower-bound calculation. Within each DPU, tasklets divide the local lists into disjoint portions and process them in parallel. For each vector $x$, the PIM-native kernel in Section~\ref{sec:design-kernel} computes $\mathrm{LB}_2(q,x)$. The DPU emits the vector identifier only if
\[
    \mathrm{LB}_2(q,x)\le\tau^2(q).
\]
These independent decisions require no query-dependent LUT, ranking buffer, or per-vector ranking synchronization.

After filtering, the CPU gathers survivors from all DPUs and reevaluates them using full multi-bit RaBitQ codes. It inserts the refined candidates into a CPU top-$k$ heap to obtain the final ordering.

Algorithm~\ref{alg:filterpim} summarizes the online procedure. Across consecutive batches, CPU-side routing, threshold estimation, cluster filtering, and reranking overlap with PIM-side vector filtering. Section~\ref{sec:implementation} describes this pipelined CPU--PIM execution in detail.

\begin{algorithm}[t]
\small
\caption{\sys{} cooperative query processing}
\label{alg:filterpim}
\begin{algorithmic}[1]

\Procedure{SearchBatch}{$Q,nprobe,k$}

    \Statex \Comment{CPU: threshold and cluster filtering}
    \State $\{J_q\}_{q\in Q}
        \gets \Call{CoarseSearch}{Q,nprobe}$

    \ForAll{$q\in Q$}
        \State $\tau_q
            \gets \Call{CalibratedThreshold}
            {q,J_q,\{H_j\},k}$
        \State $\mathcal{A}_q\gets\emptyset$

        \ForAll{$j\in J_q$}
            \State $m_j
                \gets \Call{SurvivorMass}
                {q,C_j,H_j,\tau_q}$
            \If{$m_j\ge\mu_{\min}$}
                \State $\mathcal{A}_q
                    \gets \mathcal{A}_q\cup\{j\}$
            \EndIf
        \EndFor

        \State \Call{Dispatch}{$q,\mathcal{A}_q,\tau_q^2$}
    \EndFor

    \Statex \Comment{PIM: independent vector filtering}
    \ForAll{$(q,j,\tau_q^2)$ \textbf{ dispatched in parallel}}
        \ForAll{$x_i\in C_j$}
            \State $\underline{d}_i^2
                \gets \Call{RaBitQLowerBound}{q,x_i}$
            \If{$\underline{d}_i^2\le\tau_q^2$}
                \State \Call{EmitSurvivor}{$q,i$}
            \EndIf
        \EndFor
    \EndFor

    \Statex \Comment{CPU: survivor reranking}
    \ForAll{$q\in Q$}
        \State $\mathcal{S}_q
            \gets \Call{GatherSurvivors}{q}$
        \State $R_q
            \gets \Call{Rerank}{q,\mathcal{S}_q,k}$
    \EndFor

    \State \Return $\{R_q\}_{q\in Q}$

\EndProcedure

\end{algorithmic}
\end{algorithm}

\section{Implementation}
\label{sec:implementation}

We implement \sys{} in C++ on top of FAISS's IVF--RaBitQ index~\cite{faiss} and the UPMEM SDK~\cite{upmem-doc}. We first describe the PIM data path and scheduling, followed by two execution optimizations: configuration-specialized DPU kernels and pipelined CPU--PIM execution.

\subsection{PIM Data Path and Work Scheduling}
\label{sec:impl-layout}

\sys{} adopts the list-slicing and hot-list replication strategies used by prior PIM-based ANNS systems~\cite{chen2024upanns,chen2025drimann,wu2025pimann}. Large IVF lists are divided into contiguous slices and distributed across DPUs, while frequently accessed lists are replicated so that concurrent queries can use different replicas. The CPU retains the IVF--RaBitQ index, residual histograms, and placement metadata. Each PIM-resident record stores the vector identifier, 1-bit base code, and lower-bound factors contiguously in MRAM. For each query--cluster pair that survives cluster filtering, the CPU computes the query residual and encoded query once and packages them with $\tau^2(q)$ and the auxiliary query factors into DPU work records. Each work item is assigned to a DPU containing the corresponding slice. When multiple replicas are available, \sys{} selects the resident DPU with the lowest assigned scan volume in the current batch. Within each DPU, tasklets divide the scheduled slices into disjoint portions for parallel scanning.

Each tasklet accumulates survivor identifiers in a private DMA-sized WRAM buffer. When the buffer fills, the tasklet briefly synchronizes to reserve space in a DPU-wide MRAM output region and flushes the identifiers in one DMA transfer. After filtering, the host retrieves only populated output regions through scatter--gather transfers~\cite{upmem-doc} and groups the identifiers by query for CPU-side multi-bit reranking. This keeps synchronization outside the per-vector hot path and transfers only surviving candidates.

\subsection{Configuration-Specialized DPU Kernel}
\label{sec:impl-kernel}

A generic DPU kernel must support different vector dimensions, $qb$ values, tasklet counts, and buffer configurations at runtime. This introduces configuration-dependent branches and generic loops into the scan hot path. Runtime-sized buffers also require dynamic WRAM allocation on each kernel launch, while PIM-side dynamic allocation incurs substantial overhead and tasklet contention~\cite{pim-alloc}. Conservative maximum-sized buffers avoid allocation but waste scarce WRAM and limit tasklet concurrency. Runtime-dependent loop bounds further prevent the compiler from fully optimizing and unrolling the distance-calculation path.

Because these parameters are fixed once the index and configuration are known, \sys{} generates a specialized DPU binary during index preparation. The generator specializes the vector dimension, $qb$, tasklet count, DMA size, output-buffer size, and FastFloat significand width. It statically allocates tasklet-local buffers, providing a predictable WRAM footprint and more effective WRAM utilization. It also removes configuration branches, fixes loop bounds, emits an unrolled XOR--population-count path, and inlines the fused FastFloat product into the lower-bound calculation. These choices reduce allocation and control-flow overhead while exposing a straight-line scan path that better sustains DPU pipeline utilization. Generated binaries are compiled once, cached, and reused for all searches with the same configuration.

\subsection{Pipelined CPU--PIM Execution}
\label{sec:impl-runtime}

\begin{figure}[t]
    \centering
    \includegraphics[width=\linewidth]{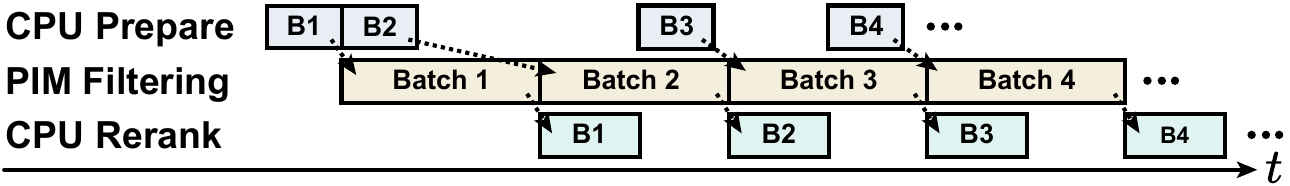}
    \vspace{-12pt}
    \caption{\textbf{Pipelined CPU--PIM execution.}}
    \label{fig:pipeline}
    \vspace{-6pt}
\end{figure}

A synchronous schedule serializes CPU preparation, PIM filtering, survivor retrieval, and CPU reranking, leaving either the CPU or DPUs idle. \sys{} instead pipelines consecutive query batches, as illustrated in Figure~\ref{fig:pipeline}. While the DPUs filter batch $i$, the CPU prepares batch $i+1$---including coarse routing, survivor-mass estimation, cluster filtering, scheduling, and query encoding---and reranks the survivors from batch $i-1$. Once filtering of batch $i$ completes, the host retrieves its survivor identifiers and immediately launches batch $i+1$. The CPU then parses and reranks batch $i$ while the DPUs filter batch $i+1$, sustaining overlap across steady-state execution.

\section{Evaluation}
\label{sec:evaluation}

\subsection{Experimental Setup}
\label{sec:eval-setup}

\paragraph{Testbed}
We conduct all experiments on a dual-socket server with two Intel Xeon Silver 4216 CPUs at 2.10\,GHz, 251\,GiB of DDR4 memory, and 20 UPMEM DIMMs containing 2,543 operational DPUs at 400\,MHz. The server runs Ubuntu 22.04.5 and UPMEM SDK 2025.1.0. Host and DPU code are compiled with GCC 11.4 and UPMEM Clang 12, respectively, using \texttt{-O3}. FAISS is built with AVX-512 and Intel MKL enabled.

\paragraph{Datasets}
Following prior PIM ANNS work~\cite{chen2025drimann}, we evaluate \sys{} on the billion-scale SIFT1B~\cite{jegou2011product} and DEEP1B~\cite{babenko2016deep} datasets. Each dataset provides 10,000 queries. We reserve 1,000 queries for calibration and use the remaining 9,000 for evaluation. We additionally use MS MARCO10M, a subset of MS MARCO~\cite{nguyen2016msmarco}, to evaluate high-dimensional performance. Parameter sweeps and hardware-scaling experiments use a 100M subset of DEEP1B, denoted DEEP100M.

\begin{table}[t]
    \centering
    \caption{\textbf{Dataset characteristics.}}
    \label{tab:datasets}
    \small
    \setlength{\tabcolsep}{4.5pt}
    \begin{tabular}{lrrrr}
        \hline
        Dataset & Vectors & Queries & Dim. & Type \\
        \hline
        DEEP1B      & 1B  & 9,000 & 96    & float32 \\
        SIFT1B      & 1B  & 9,000 & 128   & uint8 \\
        MS MARCO10M & 10M & 1,000 & 1,024 & float32 \\
        \hline
    \end{tabular}
\end{table}

\paragraph{Baselines}
\begin{itemize}
    \item \textbf{FAISS-IVF-RaBitQ} is a CPU implementation of IVF--RaBitQ by FAISS~\cite{faiss}. It uses the same index and multi-bit reranking path as \sys{}.
    \item \textbf{DRIM-ANN}~\cite{chen2025drimann} is a prior PIM-based IVFPQ system. We adapt its open-source implementation and use its original PQ-$M=16$ configuration with INT8 inputs.
    \item \textbf{DRIM-ANN-Rerank} extends DRIM-ANN with CPU-side reranking for a fairer high-recall comparison. We set each DPU's local top-$k$ capacity to 500, the largest that fits within its 64\,KB WRAM. The CPU reranks the retained candidates to produce the final top-$k$ results, with this cost included in end-to-end latency.
\end{itemize}
We compare against DRIM-ANN~\cite{chen2025drimann}, the state-of-the-art commodity-PIM ANNS system. UPVSS~\cite{liu2025upvss} cannot execute at the 1B-vector scale because its 384--512\,GB full-precision index exceeds PIM capacity. We exclude UpANNS~\cite{chen2024upanns} because it uses a similar IVFPQ design, is reported to underperform DRIM-ANN, and its per-tasklet top-$k$ exceeds DPU WRAM at $k=500$. PIMANN~\cite{wu2025pimann} focuses on optimizing per-core scheduling and is orthogonal to our work, as discussed in Section~\ref{sec:related}.

\paragraph{Metrics and methodology}
We report Recall@$k$ and end-to-end throughput in queries per second (QPS). End-to-end time includes online CPU computation, PIM execution, and CPU--PIM transfers. Index construction, calibration, kernel compilation, and initial database loading are offline and excluded. Unless otherwise stated, \sys{} uses a batch size of 1,000, $k=100$, query quantization width $qb=4$, multi-bit code width $nb=8$, $\epsilon=0.01$, $\delta=0.002$, $N_h=64$, and seven FastFloat fraction bits. \sys{} and DRIM-ANN use comparable list-slicing and replication configurations.

\subsection{End-to-End Performance}
\label{sec:eval-endtoend}

\begin{figure*}[t]
    \centering
    \includegraphics[width=\textwidth]{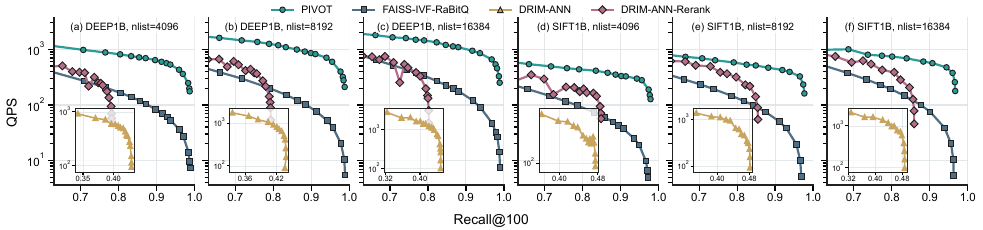}
    \vspace{-12pt}
    \caption{\textbf{End-to-end throughput--recall tradeoff.} Recall@100 and QPS on DEEP1B and SIFT1B across three $nlist$ settings. Each point varies $nprobe$.}
    \label{fig:e2e}
    \vspace{-12pt}
\end{figure*}

Figure~\ref{fig:e2e} compares the end-to-end throughput--recall tradeoff of \sys{} with the CPU and PIM baselines. Across both datasets and all $nlist$ settings, \sys{} consistently sustains higher QPS, with its advantage widening toward the high-recall region. Higher recall requires larger $nprobe$ and exposes more vector-scanning work, allowing \sys{} to better exploit PIM bandwidth and parallelism. \sys{} and FAISS-IVF-RaBitQ reach nearly identical maximum recall, confirming that the PIM-native kernel and conservative filtering preserve search accuracy.

DRIM-ANN remains in a substantially lower-recall region because its PQ-$M=16$ configuration is constrained by WRAM capacity, leading to greater quantization error. This limitation is more pronounced on DEEP1B, where DRIM-ANN additionally quantizes the original float32 vectors to INT8 for PIM execution. DRIM-ANN-Rerank improves recall, but its fixed DPU-local top-500 pool may discard globally useful candidates before CPU reranking and cannot be enlarged due to the WRAM limit. These constraints prevent prior PIM systems from reaching practical high-recall operating points, whereas \sys{} improves throughput across the curve while preserving high recall, making commodity PIM more practical for real-world ANNS.

\begin{figure}[t]
    \centering
    \includegraphics[width=\linewidth]{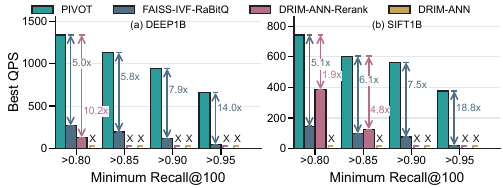}
    \vspace{-12pt}
    \caption{\textbf{Best QPS at minimum Recall@100 targets.} Each bar reports the highest measured QPS across all $nlist$ and $nprobe$ settings. A cross indicates that no measured configuration reaches the target.}
    \label{fig:fixed-recall}
    \vspace{-6pt}
\end{figure}

Figure~\ref{fig:fixed-recall} makes this trend explicit. Compared with FAISS-IVF-RaBitQ, \sys{} provides $5.0$--$5.1\times$ the QPS at Recall@100 above 0.80, with the advantage widening to $14.0$--$18.8\times$ above 0.95. At targets reached by DRIM-ANN-Rerank, \sys{} provides up to $10.2\times$ and $4.8\times$ the QPS on DEEP1B and SIFT1B, respectively. DRIM-ANN-Rerank does not reach targets above 0.85 on DEEP1B or above 0.90 on SIFT1B. These results show that combining the PIM-native distance-evaluation kernel with cooperative filtering accelerates scan-heavy settings while preserving high recall without distributed top-$k$ state.

\subsection{Effectiveness of Cooperative Filtering}
\label{sec:eval-filtering}

\begin{figure}[t]
    \centering
    \includegraphics[width=\linewidth]{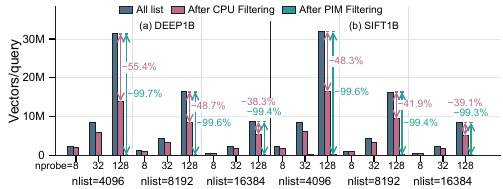}
    \vspace{-12pt}
    \caption{\textbf{Cooperative filtering work reduction.} Vectors per query from all probed lists, after CPU filtering, and after PIM filtering.}
    \label{fig:filter-work}
    \vspace{-6pt}
\end{figure}

Figure~\ref{fig:filter-work} reports the number of vectors remaining after each filtering stage. At $nprobe=128$, CPU cluster filtering removes 38.3--55.4\% of probed DEEP1B vectors and 39.1--48.3\% of probed SIFT1B vectors before dispatch. PIM vector filtering then reduces the admitted vectors to a small survivor set. Together, the two stages eliminate 99.3--99.7\% of the initially probed vectors. Cluster filtering reduces DPU scan work, while vector filtering reduces PIM--CPU transfer and CPU reranking. CPU cluster filtering is more effective at larger $nprobe$ and smaller $nlist$. Larger $nprobe$ introduces more marginal clusters, while larger $nlist$ produces finer and shorter lists with less low-yield work.

\begin{figure}[t]
    \centering
    \includegraphics[width=\linewidth]{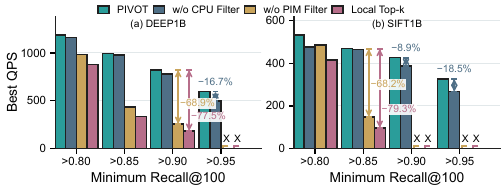}
    \vspace{-12pt}
    \caption{\textbf{Cooperative filtering ablation.} ``w/o CPU Filter'' disables CPU cluster filtering. ``w/o PIM Filter'' replaces PIM vector filtering with fixed per-DPU top-500 selection followed by CPU reranking. ``Local Top-$k$'' disables both filtering stages and uses the same selection and reranking path. A cross indicates that the recall target is not reached.}
    \label{fig:filter-ablation}
    \vspace{-6pt}
\end{figure}

Figure~\ref{fig:filter-ablation} shows how these work reductions translate into end-to-end QPS. At the highest recall target reached by all variants, replacing PIM vector filtering with DPU-local top-500 lowers QPS by 68.9\% on DEEP1B and 68.2\% on SIFT1B. Disabling both filtering stages lowers QPS by 77.5\% and 79.3\%, respectively. CPU cluster filtering contributes more as $nprobe$ grows and low-yield clusters enter the probed set. The local top-500 variants fail to reach higher recall because WRAM capacity limits each DPU's candidate pool, restricting the candidates available for CPU reranking. These results show that both filtering stages are necessary for the throughput and recall benefits of cooperative filtering.

\subsection{Kernel and Execution Optimizations}
\label{sec:eval-techniques}

\begin{figure}[t]
    \centering
    \includegraphics[width=\linewidth]{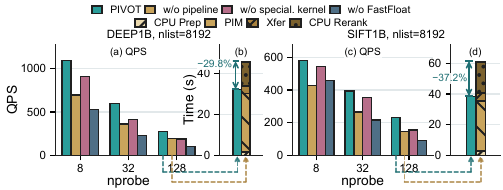}
    \vspace{-12pt}
    \caption{\textbf{Kernel and execution ablation.} QPS at $nprobe\in\{8,32,128\}$ and time breakdowns at $nprobe=128$. ``w/o pipeline'' disables pipelined CPU--PIM execution. ``w/o special. kernel'' uses a generic DPU kernel. ``w/o FastFloat'' uses UPMEM's software-emulated floating-point routines.}
    \label{fig:impl-ablation}
    \vspace{-6pt}
\end{figure}

Figure~\ref{fig:impl-ablation} evaluates FastFloat, kernel specialization, and pipelined CPU--PIM execution. FastFloat accelerates lower-bound reconstruction in the PIM-native kernel, improving QPS by 2.57--2.75$\times$ over general software-emulated floating point. Across both datasets and all evaluated $nprobe$ settings, it changes Recall@100 by at most 0.07 percentage points, indicating negligible impact on search accuracy. The gain is slightly larger on 96-dimensional DEEP1B because reconstruction accounts for a larger fraction of scan time. Pipelining reduces end-to-end time by 29.8\% on DEEP1B and 37.2\% on SIFT1B by overlapping CPU preparation and reranking with PIM filtering. Kernel specialization improves QPS by 1.49--1.50$\times$ by eliminating runtime control flow, generic loops, and variable buffer management. Together, these results validate FastFloat for the PIM-native kernel and demonstrate the end-to-end benefits of specialization and pipelining.

\subsection{Sensitivity Analysis}
\label{sec:eval-sensitivity}

We use DEEP100M with $nlist=8192$ and $nprobe=2048$ for parameter sweeps. The smaller database keeps experiments tractable, while the large $nprobe$ creates a scan-heavy regime representative of billion-scale search. These experiments characterize trends rather than absolute performance.

\begin{figure}[t]
    \centering
    \includegraphics[width=\linewidth]{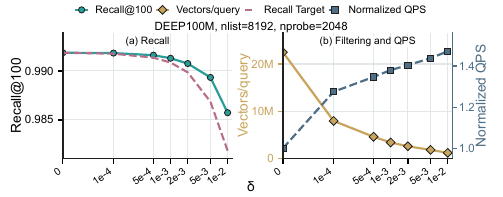}
    \vspace{-14pt}
    \caption{\textbf{CPU cluster-filtering sensitivity.} Recall@100, vectors per query, and normalized QPS as the recall-loss budget $\delta$ increases. The dashed line shows the recall target. Setting $\delta=0$ disables CPU cluster filtering.}
    \label{fig:delta}
    \vspace{-6pt}
\end{figure}

Figure~\ref{fig:delta} evaluates the recall-loss budget $\delta$. Increasing $\delta$ permits a more aggressive $\mu_{\min}$, filtering more low-yield clusters, reducing vectors scheduled for PIM, and increasing QPS. Measured recall remains above the configured target throughout the sweep, showing that survivor-mass filtering provides a predictable accuracy--throughput tradeoff.

\begin{figure}[t]
    \centering
    \includegraphics[width=\linewidth]{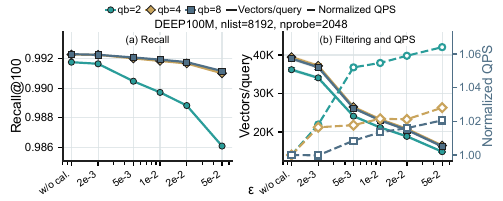}
    \vspace{-14pt}
    \caption{\textbf{Threshold-calibration sensitivity.} Recall@100, survivors per query, and normalized QPS under different calibration risks $\epsilon$ and query quantization widths $qb$. The leftmost point, ``w/o cal.,'' uses the uncalibrated threshold.}
    \label{fig:epsilon}
    \vspace{-6pt}
\end{figure}

Figure~\ref{fig:epsilon} evaluates the calibration risk $\epsilon$ and query quantization width $qb$. Increasing $\epsilon$ produces a less conservative threshold, reducing survivors and increasing QPS while gradually lowering recall. This confirms that $\epsilon$ controls the recall--reranking-work tradeoff of PIM vector filtering. The $qb=4$ and $qb=8$ curves remain close, while $qb=2$ is more sensitive to threshold calibration. Consistent with RaBitQ's observation that 4-bit query quantization introduces negligible error~\cite{gao2024rabitq}, we use $qb=4$ by default.

\begin{figure}[t]
    \centering
    \includegraphics[width=\linewidth]{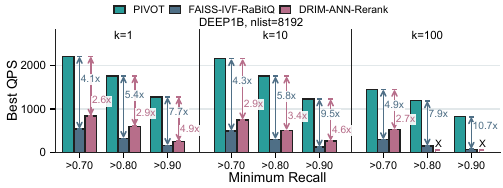}
    \vspace{-14pt}
    \caption{\textbf{Sensitivity to the top-$k$ setting.} Best QPS at minimum Recall@$k$ targets for $k\in\{1,10,100\}$.}
    \label{fig:k}
    \vspace{-6pt}
\end{figure}

Figure~\ref{fig:k} evaluates different top-$k$ settings. \sys{} consistently provides the highest QPS across different $k$, outperforming FAISS-IVF-RaBitQ by $4.1$--$10.7\times$ and DRIM-ANN-Rerank by $2.6$--$4.9\times$ at targets reached by both systems. DRIM-ANN-Rerank's recall degrades as $k$ increases. At $k=100$, it reaches only the 0.70 recall target because its fixed 500-entry DPU-local pool provides limited oversampling headroom and cannot be enlarged due to the WRAM limit. In contrast, \sys{} maintains high recall without PIM-side top-$k$ state that grows with $k$.

\subsection{Hardware and Dimensional Scalability}
\label{sec:eval-scalability}

\begin{figure}[t]
    \centering
    \includegraphics[width=\linewidth]{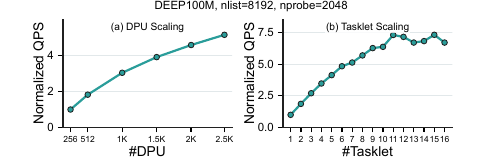}
    \vspace{-14pt}
    \caption{\textbf{Hardware scalability.} DPU and tasklet scaling on DEEP100M.}
    \label{fig:scaling}
    \vspace{-6pt}
\end{figure}

Figure~\ref{fig:scaling} evaluates DPU- and tasklet-level scaling. Throughput increases steadily with DPU count, showing that the PIM-native distance-evaluation kernel scales across distributed memory-side processors. Scaling remains sublinear because CPU-side stages do not scale with DPU count and uneven IVF-list sizes limit load balance. Tasklet throughput rises rapidly up to approximately 11 tasklets and then plateaus. Since instructions from the same tasklet are dispatched 11 cycles apart on UPMEM, approximately 11 tasklets are sufficient to fill the DPU pipeline~\cite{gomezluna2022benchmarking}. Additional tasklets provide little benefit while increasing WRAM pressure.

\begin{figure}[t]
    \centering
    \includegraphics[width=\linewidth]{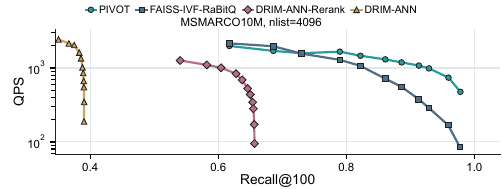}
    \vspace{-14pt}
    \caption{\textbf{High-dimensional performance.} End-to-end throughput--recall tradeoff on the 1,024-dimensional MS MARCO10M dataset.}
    \label{fig:msmarco}
    \vspace{-6pt}
\end{figure}

Figure~\ref{fig:msmarco} evaluates \sys{} on high-dimensional embeddings. At low recall, the smaller database and low $nprobe$ expose little vector-scanning work, so fixed CPU--PIM preparation and transfer costs are not fully amortized, making \sys{} slightly slower than FAISS-IVF-RaBitQ. We include this region for completeness, while practical deployments primarily target higher recall. As recall and $nprobe$ increase, vector scanning becomes dominant and \sys{} progressively outperforms the CPU baseline, reaching $4.38\times$ the QPS at Recall@100 above 0.95. The DRIM-ANN variants remain below 0.66 Recall@100 because PQ-$M=16$ provides insufficient quantization granularity for $D=1024$. Increasing $M$ could improve accuracy, but would enlarge the query-dependent LUT and further compete with local top-$k$ state for WRAM. These results show that \sys{} remains effective as vector dimensionality increases.

Overall, the results validate \sys{}'s design. The PIM-native distance-evaluation kernel and threshold-guided cooperative filtering preserve high recall while avoiding distributed top-$k$ state. Configuration-specialized kernels and pipelined CPU--PIM execution provide scalable end-to-end performance.

\section{Related Work}
\label{sec:related}

\textbf{ANNS on PIM and near-data accelerators:}
UpANNS~\cite{chen2024upanns} and DRIM-ANN~\cite{chen2025drimann} map IVFPQ search to commercial UPMEM hardware through PIM-aware data placement, scheduling, and resource management. UpANNS maintains per-tasklet candidate heaps, while DRIM-ANN uses a shared DPU-local top-$k$ queue. Both retain distributed ranking state and query-dependent PQ lookup tables. UPVSS~\cite{liu2025upvss} performs floating-point distance evaluation on DPUs and returns all distances and identifiers for CPU-side selection, avoiding PIM-side ranking but incurring substantial output traffic. PIMANN~\cite{wu2025pimann} improves utilization through persistent kernels and fine-grained per-core scheduling. Its scheduling mechanisms are orthogonal to \sys{} and could complement our design. Other work uses customized near-memory, near-storage, or in-memory hardware~\cite{zhu2023pyramid,li2025ansmet,liu2025seim,zou2026naszip}. These designs require specialized distance units, filtering logic, or data paths unavailable on commodity PIM. In contrast, \sys{} runs on unmodified UPMEM hardware and replaces distributed ranking with a CPU-derived radius for CPU-side cluster filtering and independent DPU-side lower-bound filtering.

\textbf{ANNS on CPUs, GPUs, and FPGAs:}
CPU systems optimize quantization- and graph-based ANNS through SIMD execution, multithreading, cache-aware layouts, and storage-aware indexes. Representative systems include FAISS and ScaNN for quantization-based search, HNSW and NSG for in-memory graph traversal, and DiskANN and SPANN for billion-scale search across memory and SSDs~\cite{johnson2019billion,guo2020scann,malkov2020hnsw,fu2019nsg,subramanya2019diskann,chen2021spann}. GPU systems exploit thread parallelism and device-memory bandwidth for IVF and graph search~\cite{zhao2020song,ootomo2024cagra,khan2024bang,liu2024juno}. FPGA and dedicated accelerators customize distance units, pipelines, and memory access for higher throughput and energy efficiency~\cite{zhang2018fpga,lee2022anna,zeng2023dfgas,yuan2025fanns}. These platforms offer different architectural advantages and operate under different constraints, and thus are not directly comparable to programmable commodity PIM.

\textbf{Systems on commodity PIM:}
Prior work characterizes UPMEM's compute, memory, communication, and scaling behavior~\cite{gomezluna2022benchmarking,hyun2024pathfinding}. Architecture-aware systems redesign data structures and runtimes for these constraints. PIM-tree and PIM-trie use CPU--PIM work partitioning, data placement, and skew-aware scheduling for ordered indexes~\cite{kang2022pimtree,kang2023pimtrie}. SimplePIM provides reusable programming and communication primitives, while PID-Comm optimizes collective communication across isolated DPUs~\cite{chen2023simplepim,noh2024pidcomm}. Other work adapts numerical workloads to the MRAM--WRAM hierarchy, studies architecture-aware algorithm mapping, and reduces dynamic-allocation overhead~\cite{diab2023aim,rhyner2024pimopt,pim-alloc}. These efforts establish the importance of work partitioning, communication reduction, WRAM management, and PIM-friendly operations. \sys{} applies these principles to redesign ANNS distance evaluation and candidate selection.

\section{Conclusion}

This paper rethinks ANNS execution on commodity PIM by addressing both the evaluation--kernel and search--paradigm mismatches. \sys{} combines a PIM-native distance-evaluation kernel with threshold-guided cooperative filtering. The CPU derives a conservative query-specific radius and filters low-yield clusters, while the DPUs independently filter vectors and return only survivor identifiers for CPU reranking. Configuration-specialized kernels and pipelined CPU--PIM execution sustain high utilization across both sides. On real UPMEM hardware, \sys{} filters out 99.3--99.7\% of probed vectors and provides up to $18.8\times$ and $10.2\times$ the QPS of CPU and prior PIM baselines, respectively. It also preserves high recall across different top-$k$ settings and remains effective on high-dimensional embeddings. These results show that efficient PIM-based ANNS requires jointly redesigning the distance-evaluation kernel and candidate-selection paradigm for the underlying architecture, rather than directly transplanting CPU-oriented workflows.


\bibliographystyle{IEEEtranS}
\bibliography{refs}

@inproceedings{lewis2020rag,
 author = {Lewis, Patrick and Perez, Ethan and Piktus, Aleksandra and Petroni, Fabio and Karpukhin, Vladimir and Goyal, Naman and K\"{u}ttler, Heinrich and Lewis, Mike and Yih, Wen-tau and Rockt\"{a}schel, Tim and Riedel, Sebastian and Kiela, Douwe},
 booktitle = {Advances in Neural Information Processing Systems},
 editor = {H. Larochelle and M. Ranzato and R. Hadsell and M.F. Balcan and H. Lin},
 pages = {9459--9474},
 publisher = {Curran Associates, Inc.},
 title = {Retrieval-Augmented Generation for Knowledge-Intensive NLP Tasks},
 url = {https://proceedings.neurips.cc/paper_files/paper/2020/file/6b493230205f780e1bc26945df7481e5-Paper.pdf},
 volume = {33},
 year = {2020}
}

@InProceedings{guu2020realm,
  title = 	 {Retrieval Augmented Language Model Pre-Training},
  author =       {Guu, Kelvin and Lee, Kenton and Tung, Zora and Pasupat, Panupong and Chang, Mingwei},
  booktitle = 	 {Proceedings of the 37th International Conference on Machine Learning},
  pages = 	 {3929--3938},
  year = 	 {2020},
  editor = 	 {III, Hal Daumé and Singh, Aarti},
  volume = 	 {119},
  series = 	 {Proceedings of Machine Learning Research},
  month = 	 {13--18 Jul},
  publisher =    {PMLR},
  url = 	 {https://proceedings.mlr.press/v119/guu20a.html}
}

@inproceedings{karpukhin2020dpr,
    title = "Dense Passage Retrieval for Open-Domain Question Answering",
    author = "Karpukhin, Vladimir  and
      Oguz, Barlas  and
      Min, Sewon  and
      Lewis, Patrick  and
      Wu, Ledell  and
      Edunov, Sergey  and
      Chen, Danqi  and
      Yih, Wen-tau",
    editor = "Webber, Bonnie  and
      Cohn, Trevor  and
      He, Yulan  and
      Liu, Yang",
    booktitle = "Proceedings of the 2020 Conference on Empirical Methods in Natural Language Processing (EMNLP)",
    month = nov,
    year = "2020",
    address = "Online",
    publisher = "Association for Computational Linguistics",
    url = "https://aclanthology.org/2020.emnlp-main.550/",
    doi = "10.18653/v1/2020.emnlp-main.550",
    pages = "6769--6781"
}

@inproceedings{wang2021milvus,
author = {Wang, Jianguo and Yi, Xiaomeng and Guo, Rentong and Jin, Hai and Xu, Peng and Li, Shengjun and Wang, Xiangyu and Guo, Xiangzhou and Li, Chengming and Xu, Xiaohai and Yu, Kun and Yuan, Yuxing and Zou, Yinghao and Long, Jiquan and Cai, Yudong and Li, Zhenxiang and Zhang, Zhifeng and Mo, Yihua and Gu, Jun and Jiang, Ruiyi and Wei, Yi and Xie, Charles},
title = {Milvus: A Purpose-Built Vector Data Management System},
year = {2021},
isbn = {9781450383431},
publisher = {Association for Computing Machinery},
address = {New York, NY, USA},
url = {https://doi.org/10.1145/3448016.3457550},
doi = {10.1145/3448016.3457550},
booktitle = {Proceedings of the 2021 International Conference on Management of Data},
pages = {2614–2627},
numpages = {14},
location = {Virtual Event, China},
series = {SIGMOD '21}
}

@article{guo2022manu,
author = {Guo, Rentong and Luan, Xiaofan and Xiang, Long and Yan, Xiao and Yi, Xiaomeng and Luo, Jigao and Cheng, Qianya and Xu, Weizhi and Luo, Jiarui and Liu, Frank and Cao, Zhenshan and Qiao, Yanliang and Wang, Ting and Tang, Bo and Xie, Charles},
title = {Manu: a cloud native vector database management system},
year = {2022},
issue_date = {August 2022},
publisher = {VLDB Endowment},
volume = {15},
number = {12},
issn = {2150-8097},
url = {https://doi.org/10.14778/3554821.3554843},
doi = {10.14778/3554821.3554843},
journal = {Proc. VLDB Endow.},
month = aug,
pages = {3548–3561},
numpages = {14}
}

@inproceedings {zhang2023vbase,
author = {Qianxi Zhang and Shuotao Xu and Qi Chen and Guoxin Sui and Jiadong Xie and Zhizhen Cai and Yaoqi Chen and Yinxuan He and Yuqing Yang and Fan Yang and Mao Yang and Lidong Zhou},
title = {{VBASE}: Unifying Online Vector Similarity Search and Relational Queries via Relaxed Monotonicity},
booktitle = {17th USENIX Symposium on Operating Systems Design and Implementation (OSDI 23)},
year = {2023},
isbn = {978-1-939133-34-2},
address = {Boston, MA},
pages = {377--395},
url = {https://www.usenix.org/conference/osdi23/presentation/zhang-qianxi},
publisher = {USENIX Association},
month = jul
}

@inproceedings{covington2016youtube,
author = {Covington, Paul and Adams, Jay and Sargin, Emre},
title = {Deep Neural Networks for YouTube Recommendations},
year = {2016},
isbn = {9781450340359},
publisher = {Association for Computing Machinery},
address = {New York, NY, USA},
url = {https://doi.org/10.1145/2959100.2959190},
doi = {10.1145/2959100.2959190},
booktitle = {Proceedings of the 10th ACM Conference on Recommender Systems},
pages = {191–198},
numpages = {8},
location = {Boston, Massachusetts, USA},
series = {RecSys '16}
}

@inproceedings{yi2019sampling,
author = {Yi, Xinyang and Yang, Ji and Hong, Lichan and Cheng, Derek Zhiyuan and Heldt, Lukasz and Kumthekar, Aditee and Zhao, Zhe and Wei, Li and Chi, Ed},
title = {Sampling-bias-corrected neural modeling for large corpus item recommendations},
year = {2019},
isbn = {9781450362436},
publisher = {Association for Computing Machinery},
address = {New York, NY, USA},
url = {https://doi.org/10.1145/3298689.3346996},
doi = {10.1145/3298689.3346996},
booktitle = {Proceedings of the 13th ACM Conference on Recommender Systems},
pages = {269–277},
numpages = {9},
location = {Copenhagen, Denmark},
series = {RecSys '19}
}

@inproceedings{xiong2021ance,
author = {Xiong, Lee and Xiong, Chenyan and Li, Ye and Tang, Kwok-Fung and Liu, Jialin and Bennett, Paul and Ahmed, Junaid and Overwijk, Arnold},
title = {Approximate Nearest Neighbor Negative Contrastive Learning for Dense Text Retrieval},
booktitle = {International Conference on Learning Representations (ICLR)},
year = {2021},
month = {April},
url = {https://www.microsoft.com/en-us/research/publication/approximate-nearest-neighbor-negative-contrastive-learning-for-dense-text-retrieval/},
}

@inproceedings{reimers2019sbert,
    title = "Sentence-{BERT}: Sentence Embeddings using {S}iamese {BERT}-Networks",
    author = "Reimers, Nils  and
      Gurevych, Iryna",
    editor = "Inui, Kentaro  and
      Jiang, Jing  and
      Ng, Vincent  and
      Wan, Xiaojun",
    booktitle = "Proceedings of the 2019 Conference on Empirical Methods in Natural Language Processing and the 9th International Joint Conference on Natural Language Processing (EMNLP-IJCNLP)",
    month = nov,
    year = "2019",
    address = "Hong Kong, China",
    publisher = "Association for Computational Linguistics",
    url = "https://aclanthology.org/D19-1410/",
    doi = "10.18653/v1/D19-1410",
    pages = "3982--3992"
}

@inproceedings{santhanam2022colbertv2,
    title = "{C}ol{BERT}v2: Effective and Efficient Retrieval via Lightweight Late Interaction",
    author = "Santhanam, Keshav  and
      Khattab, Omar  and
      Saad-Falcon, Jon  and
      Potts, Christopher  and
      Zaharia, Matei",
    editor = "Carpuat, Marine  and
      de Marneffe, Marie-Catherine  and
      Meza Ruiz, Ivan Vladimir",
    booktitle = "Proceedings of the 2022 Conference of the North American Chapter of the Association for Computational Linguistics: Human Language Technologies",
    month = jul,
    year = "2022",
    address = "Seattle, United States",
    publisher = "Association for Computational Linguistics",
    url = "https://aclanthology.org/2022.naacl-main.272/",
    doi = "10.18653/v1/2022.naacl-main.272",
    pages = "3715--3734"
}

@inproceedings{andoni2018annsurvey,
  title={Approximate nearest neighbor search in high dimensions},
  author={Andoni, Alexandr and Indyk, Piotr and Razenshteyn, Ilya},
  booktitle={Proceedings of the International Congress of Mathematicians: Rio de Janeiro 2018},
  pages={3287--3318},
  year={2018},
  organization={World Scientific}
}

@article{wang2021graphsurvey,
author = {Wang, Mengzhao and Xu, Xiaoliang and Yue, Qiang and Wang, Yuxiang},
title = {A comprehensive survey and experimental comparison of graph-based approximate nearest neighbor search},
year = {2021},
issue_date = {July 2021},
publisher = {VLDB Endowment},
volume = {14},
number = {11},
issn = {2150-8097},
url = {https://doi.org/10.14778/3476249.3476255},
doi = {10.14778/3476249.3476255},
journal = {Proc. VLDB Endow.},
month = jul,
pages = {1964–1978},
numpages = {15}
}

@ARTICLE{cai2021hashing,
  author={Cai, Deng},
  journal={IEEE Transactions on Knowledge and Data Engineering}, 
  title={A Revisit of Hashing Algorithms for Approximate Nearest Neighbor Search}, 
  year={2021},
  volume={33},
  number={6},
  pages={2337-2348},
  doi={10.1109/TKDE.2019.2953897}}

@ARTICLE{jegou2011product,
  author={Jégou, Herve and Douze, Matthijs and Schmid, Cordelia},
  journal={IEEE Transactions on Pattern Analysis and Machine Intelligence}, 
  title={Product Quantization for Nearest Neighbor Search}, 
  year={2011},
  volume={33},
  number={1},
  pages={117-128},
  doi={10.1109/TPAMI.2010.57}}

@ARTICLE{johnson2019billion,
  author={Johnson, Jeff and Douze, Matthijs and Jégou, Hervé},
  journal={IEEE Transactions on Big Data}, 
  title={Billion-Scale Similarity Search with GPUs}, 
  year={2021},
  volume={7},
  number={3},
  pages={535-547},
  doi={10.1109/TBDATA.2019.2921572}}

@article{gao2024rabitq,
author = {Gao, Jianyang and Long, Cheng},
title = {RaBitQ: Quantizing High-Dimensional Vectors with a Theoretical Error Bound for Approximate Nearest Neighbor Search},
year = {2024},
issue_date = {June 2024},
publisher = {Association for Computing Machinery},
address = {New York, NY, USA},
volume = {2},
number = {3},
url = {https://doi.org/10.1145/3654970},
doi = {10.1145/3654970},
journal = {Proc. ACM Manag. Data},
month = may,
articleno = {167},
numpages = {27}
}

@article{gao2025extrabitq,
author = {Gao, Jianyang and Gou, Yutong and Xu, Yuexuan and Yang, Yongyi and Long, Cheng and Wong, Raymond Chi-Wing},
title = {Practical and Asymptotically Optimal Quantization of High-Dimensional Vectors in Euclidean Space for Approximate Nearest Neighbor Search},
year = {2025},
issue_date = {June 2025},
publisher = {Association for Computing Machinery},
address = {New York, NY, USA},
volume = {3},
number = {3},
url = {https://doi.org/10.1145/3725413},
doi = {10.1145/3725413},
journal = {Proc. ACM Manag. Data},
month = jun,
articleno = {202},
numpages = {26}
}

@INPROCEEDINGS{babenko2016deep,
  author={Babenko, Artem and Lempitsky, Victor},
  booktitle={2016 IEEE Conference on Computer Vision and Pattern Recognition (CVPR)}, 
  title={Efficient Indexing of Billion-Scale Datasets of Deep Descriptors}, 
  year={2016},
  volume={},
  number={},
  pages={2055-2063},
  doi={10.1109/CVPR.2016.226}}

@misc{nguyen2016msmarco,
      title={MS MARCO: A Human Generated MAchine Reading COmprehension Dataset}, 
      author={Payal Bajaj and Daniel Campos and Nick Craswell and Li Deng and Jianfeng Gao and Xiaodong Liu and Rangan Majumder and Andrew McNamara and Bhaskar Mitra and Tri Nguyen and Mir Rosenberg and Xia Song and Alina Stoica and Saurabh Tiwary and Tong Wang},
      year={2018},
      eprint={1611.09268},
      archivePrefix={arXiv},
      primaryClass={cs.CL},
      url={https://arxiv.org/abs/1611.09268}, 
}

@article{stam1982sphere, title={Limit theorems for uniform distributions on spheres in high-dimensional euclidean spaces}, volume={19}, DOI={10.2307/3213932}, number={1}, journal={Journal of Applied Probability}, author={Stam, A. J.}, year={1982}, pages={221–228}}

@article{wulf1995memory,
author = {Wulf, Wm. A. and McKee, Sally A.},
title = {Hitting the memory wall: implications of the obvious},
year = {1995},
issue_date = {March 1995},
publisher = {Association for Computing Machinery},
address = {New York, NY, USA},
volume = {23},
number = {1},
issn = {0163-5964},
url = {https://doi.org/10.1145/216585.216588},
doi = {10.1145/216585.216588},
journal = {SIGARCH Comput. Archit. News},
month = mar,
pages = {20–24},
numpages = {5}
}

@misc{ghose2018enabling,
      title={Enabling the Adoption of Processing-in-Memory: Challenges, Mechanisms, Future Research Directions}, 
      author={Saugata Ghose and Kevin Hsieh and Amirali Boroumand and Rachata Ausavarungnirun and Onur Mutlu},
      year={2018},
      eprint={1802.00320},
      archivePrefix={arXiv},
      primaryClass={cs.AR},
      url={https://arxiv.org/abs/1802.00320}, 
}

@article{mutlu2019processing,
title = {Processing data where it makes sense: Enabling in-memory computation},
journal = {Microprocessors and Microsystems},
volume = {67},
pages = {28-41},
year = {2019},
issn = {0141-9331},
doi = {https://doi.org/10.1016/j.micpro.2019.01.009},
url = {https://www.sciencedirect.com/science/article/pii/S0141933118302291},
author = {Onur Mutlu and Saugata Ghose and Juan Gómez-Luna and Rachata Ausavarungnirun}
}

@Inbook{mutlu2022primer,
author="Mutlu, Onur
and Ghose, Saugata
and G{\'o}mez-Luna, Juan
and Ausavarungnirun, Rachata",
editor="Aly, Mohamed M. Sabry
and Chattopadhyay, Anupam",
title="A Modern Primer on Processing in Memory",
bookTitle="Emerging Computing: From Devices to Systems: Looking Beyond Moore and Von Neumann",
year="2023",
publisher="Springer Nature Singapore",
address="Singapore",
pages="171--243",
isbn="978-981-16-7487-7",
doi="10.1007/978-981-16-7487-7_7",
url="https://doi.org/10.1007/978-981-16-7487-7_7"
}

@INPROCEEDINGS{devaux2019truepim,
  author={Devaux, Fabrice},
  booktitle={2019 IEEE Hot Chips 31 Symposium (HCS)}, 
  title={The true Processing In Memory accelerator}, 
  year={2019},
  volume={},
  number={},
  pages={1-24},
  doi={10.1109/HOTCHIPS.2019.8875680}}

@ARTICLE{gomezluna2022benchmarking,
  author={Gómez-Luna, Juan and Hajj, Izzat El and Fernandez, Ivan and Giannoula, Christina and Oliveira, Geraldo F. and Mutlu, Onur},
  journal={IEEE Access}, 
  title={Benchmarking a New Paradigm: Experimental Analysis and Characterization of a Real Processing-in-Memory System}, 
  year={2022},
  volume={10},
  number={},
  pages={52565-52608},
  doi={10.1109/ACCESS.2022.3174101}}

@INPROCEEDINGS{lee2021hbmpim,
  author={Kwon, Young-Cheon and Lee, Suk Han and Lee, Jaehoon and Kwon, Sang-Hyuk and Ryu, Je Min and Son, Jong-Pil and Seongil, O and Yu, Hak-Soo and Lee, Haesuk and Kim, Soo Young and Cho, Youngmin and Kim, Jin Guk and Choi, Jongyoon and Shin, Hyun-Sung and Kim, Jin and Phuah, BengSeng and Kim, HyoungMin and Song, Myeong Jun and Choi, Ahn and Kim, Daeho and Kim, SooYoung and Kim, Eun-Bong and Wang, David and Kang, Shinhaeng and Ro, Yuhwan and Seo, Seungwoo and Song, JoonHo and Youn, Jaeyoun and Sohn, Kyomin and Kim, Nam Sung},
  booktitle={2021 IEEE International Solid-State Circuits Conference (ISSCC)}, 
  title={25.4 A 20nm 6GB Function-In-Memory DRAM, Based on HBM2 with a 1.2TFLOPS Programmable Computing Unit Using Bank-Level Parallelism, for Machine Learning Applications}, 
  year={2021},
  volume={64},
  number={},
  pages={350-352},
  doi={10.1109/ISSCC42613.2021.9365862}}

@INPROCEEDINGS{kwon2022gddr6aim,
  author={Kwon, Yongkee and Vladimir, Kornijcuk and Kim, Nahsung and Shin, Woojae and Won, Jongsoon and Lee, Minkyu and Joo, Hyunha and Choi, Haerang and Kim, Guhyun and An, Byeongju and Kim, Jeongbin and Lee, Jaewook and Kim, Ilkon and Park, Jaehan and Park, Chanwook and Song, Yosub and Yang, Byeongsu and Lee, Hyungdeok and Kim, Seho and Kwon, Daehan and Lee, Seongju and Kim, Kyuyoung and Oh, Sanghoon and Park, Joonhong and Hong, Gimoon and Ka, Dongyoon and Hwang, Kyudong and Park, Jeongje and Kang, Kyeongpil and Kim, Jungyeon and Jeon, Junyeol and Lee, Myeongjun and Shin, Minyoung and Shin, Minhwan and Cha, Jaekyung and Jung, Changson and Chang, Kijoon and Jeong, Chunseok and Lim, Euicheol and Park, Il and Chun, Junhyun and Hynix, Sk},
  booktitle={2022 IEEE Hot Chips 34 Symposium (HCS)}, 
  title={System Architecture and Software Stack for GDDR6-AiM}, 
  year={2022},
  volume={},
  number={},
  pages={1-25},
  doi={10.1109/HCS55958.2022.9895629}}

@INPROCEEDINGS{hyun2024pathfinding,
  author={Hyun, Bongjoon and Kim, Taehun and Lee, Dongjae and Rhu, Minsoo},
  booktitle={2024 IEEE International Symposium on High-Performance Computer Architecture (HPCA)}, 
  title={Pathfinding Future PIM Architectures by Demystifying a Commercial PIM Technology}, 
  year={2024},
  volume={},
  number={},
  pages={263-279},
  doi={10.1109/HPCA57654.2024.00029}}

@article{diab2023aim,
    author = {Diab, Safaa and Nassereldine, Amir and Alser, Mohammed and Gómez Luna, Juan and Mutlu, Onur and El Hajj, Izzat},
    title = {A framework for high-throughput sequence alignment using real processing-in-memory systems},
    journal = {Bioinformatics},
    volume = {39},
    number = {5},
    pages = {btad155},
    year = {2023},
    month = {05},
    issn = {1367-4811},
    doi = {10.1093/bioinformatics/btad155},
    url = {https://doi.org/10.1093/bioinformatics/btad155},
    eprint = {https://academic.oup.com/bioinformatics/article-pdf/39/5/btad155/50204881/btad155.pdf},
}

@inproceedings{rhyner2024pimopt,
author = {Rhyner, Steve and Luo, Haocong and G\'{o}mez-Luna, Juan and Sadrosadati, Mohammad and Jiang, Jiawei and Olgun, Ataberk and Gupta, Harshita and Zhang, Ce and Mutlu, Onur},
title = {PIM-Opt: Demystifying Distributed Optimization Algorithms on a Real-World Processing-In-Memory System},
year = {2024},
isbn = {9798400706318},
publisher = {Association for Computing Machinery},
address = {New York, NY, USA},
url = {https://doi.org/10.1145/3656019.3676947},
doi = {10.1145/3656019.3676947},
booktitle = {Proceedings of the 2024 International Conference on Parallel Architectures and Compilation Techniques},
pages = {201–218},
numpages = {18},
location = {Long Beach, CA, USA},
series = {PACT '24}
}

@article{kang2022pimtree,
author = {Kang, Hongbo and Zhao, Yiwei and Blelloch, Guy E. and Dhulipala, Laxman and Gu, Yan and McGuffey, Charles and Gibbons, Phillip B.},
title = {PIM-Tree: A Skew-Resistant Index for Processing-in-Memory},
year = {2022},
issue_date = {December 2022},
publisher = {VLDB Endowment},
volume = {16},
number = {4},
issn = {2150-8097},
url = {https://doi.org/10.14778/3574245.3574275},
doi = {10.14778/3574245.3574275},
journal = {Proc. VLDB Endow.},
month = dec,
pages = {946–958},
numpages = {13}
}

@inproceedings{kang2023pimtrie,
author = {Kang, Hongbo and Zhao, Yiwei and Blelloch, Guy E. and Dhulipala, Laxman and Gu, Yan and McGuffey, Charles and Gibbons, Phillip B.},
title = {PIM-trie: A Skew-resistant Trie for Processing-in-Memory},
year = {2023},
isbn = {9781450395458},
publisher = {Association for Computing Machinery},
address = {New York, NY, USA},
url = {https://doi.org/10.1145/3558481.3591070},
doi = {10.1145/3558481.3591070},
booktitle = {Proceedings of the 35th ACM Symposium on Parallelism in Algorithms and Architectures},
pages = {1–14},
numpages = {14},
location = {Orlando, FL, USA},
series = {SPAA '23}
}

@INPROCEEDINGS{chen2023simplepim,
  author={Chen, Jinfan and Gómez-Luna, Juan and El Hajj, Izzat and Guo, Yuxin and Mutlu, Onur},
  booktitle={2023 32nd International Conference on Parallel Architectures and Compilation Techniques (PACT)}, 
  title={SimplePIM: A Software Framework for Productive and Efficient Processing-in-Memory}, 
  year={2023},
  volume={},
  number={},
  pages={99-111},
  doi={10.1109/PACT58117.2023.00017}}

@INPROCEEDINGS{noh2024pidcomm,
  author={Noh, Si Ung and Hong, Junguk and Lim, Chaemin and Park, Seongyeon and Kim, Jeehyun and Kim, Hanjun and Kim, Youngsok and Lee, Jinho},
  booktitle={2024 ACM/IEEE 51st Annual International Symposium on Computer Architecture (ISCA)}, 
  title={PID-Comm: A Fast and Flexible Collective Communication Framework for Commodity Processing-in-DIMM Devices}, 
  year={2024},
  volume={},
  number={},
  pages={245-260},
  doi={10.1109/ISCA59077.2024.00027}}

@INPROCEEDINGS{pim-alloc,
  author={Lee, Dongjae and Hyun, Bongjoon and Kwon, Youngjin and Rhu, Minsoo},
  booktitle={2026 IEEE International Symposium on High Performance Computer Architecture (HPCA)}, 
  title={PIM-Malloc: A Fast and Scalable Dynamic Memory Allocator for Processing-In-Memory (PIM) Architectures}, 
  year={2026},
  volume={},
  number={},
  pages={1-17},
  doi={10.1109/HPCA68181.2026.11408588}}

@inproceedings{chen2024upanns,
author = {Chen, Sitian and Zhou, Amelie Chi and Shi, Yucheng and Li, Yusen and Yao, Xin},
title = {UpANNS: Enhancing Billion-Scale ANNS Efficiency with Real-World PIM Architecture},
year = {2025},
isbn = {9798400714665},
publisher = {Association for Computing Machinery},
address = {New York, NY, USA},
url = {https://doi.org/10.1145/3712285.3759777},
doi = {10.1145/3712285.3759777},
booktitle = {Proceedings of the International Conference for High Performance Computing, Networking, Storage and Analysis},
pages = {789–804},
numpages = {16},
location = {
},
series = {SC '25}
}

@inproceedings{chen2025drimann,
author = {Chen, Mingkai and Han, Tianhua and Liu, Cheng and Liang, Shengwen and Yu, Kuai and Dai, Lei and Yuan, Ziming and Wang, Ying and Zhang, Lei and Li, Huawei and Li, Xiaowei},
title = {DRIM-ANN: An Approximate Nearest Neighbor Search Engine based on Commercial DRAM-PIMs},
year = {2025},
isbn = {9798400714665},
publisher = {Association for Computing Machinery},
address = {New York, NY, USA},
url = {https://doi.org/10.1145/3712285.3759801},
doi = {10.1145/3712285.3759801},
booktitle = {Proceedings of the International Conference for High Performance Computing, Networking, Storage and Analysis},
pages = {820–836},
numpages = {17},
location = {
},
series = {SC '25}
}

@inproceedings{wu2025pimann,
author = {Puqing Wu and Minhui Xie and Enrui Zhao and Dafang Zhang and Jing Wang and Xiao Liang and Kai Ren and Yunpeng Chai},
title = {Turbocharge {ANNS} on Real {Processing-in-Memory} by Enabling {Fine-Grained} {Per-PIM-Core} Scheduling},
booktitle = {2025 USENIX Annual Technical Conference (USENIX ATC 25)},
year = {2025},
isbn = {978-1-939133-48-9},
address = {Boston, MA},
pages = {1223--1241},
url = {https://www.usenix.org/conference/atc25/presentation/wu-puqing},
publisher = {USENIX Association},
month = jul
}

@INPROCEEDINGS{liu2025upvss,
  author={Liu, Chun-Chien and Wu, Chun-Feng and Jin, Yunho},
  booktitle={2025 62nd ACM/IEEE Design Automation Conference (DAC)}, 
  title={UPVSS: Jointly Managing Vector Similarity Search with Near-Memory Processing Systems}, 
  year={2025},
  volume={},
  number={},
  pages={1-7},
  doi={10.1109/DAC63849.2025.11132577}}

@INPROCEEDINGS{zhu2023pyramid,
  author={Zhu, Zhenhua and Liu, Jun and Dai, Guohao and Zeng, Shulin and Li, Bing and Yang, Huazhong and Wang, Yu},
  booktitle={2023 60th ACM/IEEE Design Automation Conference (DAC)}, 
  title={Processing-In-Hierarchical-Memory Architecture for Billion-Scale Approximate Nearest Neighbor Search}, 
  year={2023},
  volume={},
  number={},
  pages={1-6},
  doi={10.1109/DAC56929.2023.10247946}}

@inproceedings{li2025ansmet,
author = {Li, Yiwei and Jin, Yuxin and Tian, Boyu and Zhang, Huanchen and Gao, Mingyu},
title = {ANSMET: Approximate Nearest Neighbor Search with Near-Memory Processing and Hybrid Early Termination},
year = {2025},
isbn = {9798400712616},
publisher = {Association for Computing Machinery},
address = {New York, NY, USA},
url = {https://doi.org/10.1145/3695053.3731013},
doi = {10.1145/3695053.3731013},
booktitle = {Proceedings of the 52nd Annual International Symposium on Computer Architecture},
pages = {1093–1107},
numpages = {15},
location = {
},
series = {ISCA '25}
}

@INPROCEEDINGS{liu2025seim,
  author={Liu, Chaoqiang and Chen, Dan and Huang, Yu and Xiao, Wenjing and Liu, Haifeng and Zhang, Yi and Li, Huize and Liao, Xiaofei and Jin, Hai},
  booktitle={2025 62nd ACM/IEEE Design Automation Conference (DAC)}, 
  title={SeIM: In-Memory Acceleration for Approximate Nearest Neighbor Search}, 
  year={2025},
  volume={},
  number={},
  pages={1-7},
  doi={10.1109/DAC63849.2025.11132620}}

@misc{zou2026naszip,
      title={NasZip: Software and Hardware Co-Design to Accelerate Approximate Nearest Neighbor Search with DIMM-Based Near-Data Processing}, 
      author={Cheng Zou and Shuo Yang and Chen Nie and Yu Zou and Yu He and Chao Jiang and Limin Xiao and Weifeng Zhang and Zhezhi He},
      year={2026},
      eprint={2605.21952},
      archivePrefix={arXiv},
      primaryClass={cs.AR},
      url={https://arxiv.org/abs/2605.21952}, 
}

@InProceedings{guo2020scann,
  title = 	 {Accelerating Large-Scale Inference with Anisotropic Vector Quantization},
  author =       {Guo, Ruiqi and Sun, Philip and Lindgren, Erik and Geng, Quan and Simcha, David and Chern, Felix and Kumar, Sanjiv},
  booktitle = 	 {Proceedings of the 37th International Conference on Machine Learning},
  pages = 	 {3887--3896},
  year = 	 {2020},
  editor = 	 {III, Hal Daumé and Singh, Aarti},
  volume = 	 {119},
  series = 	 {Proceedings of Machine Learning Research},
  month = 	 {13--18 Jul},
  publisher =    {PMLR},
  url = 	 {https://proceedings.mlr.press/v119/guo20h.html}
}

@ARTICLE{malkov2020hnsw,
  author={Malkov, Yu A. and Yashunin, D. A.},
  journal={IEEE Transactions on Pattern Analysis and Machine Intelligence}, 
  title={Efficient and Robust Approximate Nearest Neighbor Search Using Hierarchical Navigable Small World Graphs}, 
  year={2020},
  volume={42},
  number={4},
  pages={824-836},
  doi={10.1109/TPAMI.2018.2889473}}

@article{fu2019nsg,
author = {Fu, Cong and Xiang, Chao and Wang, Changxu and Cai, Deng},
title = {Fast approximate nearest neighbor search with the navigating spreading-out graph},
year = {2019},
issue_date = {January 2019},
publisher = {VLDB Endowment},
volume = {12},
number = {5},
issn = {2150-8097},
url = {https://doi.org/10.14778/3303753.3303754},
doi = {10.14778/3303753.3303754},
journal = {Proc. VLDB Endow.},
month = jan,
pages = {461–474},
numpages = {14}
}

@inproceedings{subramanya2019diskann,
 author = {Jayaram Subramanya, Suhas and Devvrit, Fnu and Simhadri, Harsha Vardhan and Krishnawamy, Ravishankar and Kadekodi, Rohan},
 booktitle = {Advances in Neural Information Processing Systems},
 editor = {H. Wallach and H. Larochelle and A. Beygelzimer and F. d\textquotesingle Alch\'{e}-Buc and E. Fox and R. Garnett},
 pages = {},
 publisher = {Curran Associates, Inc.},
 title = {DiskANN: Fast Accurate Billion-point Nearest Neighbor Search on a Single Node},
 url = {https://proceedings.neurips.cc/paper_files/paper/2019/file/09853c7fb1d3f8ee67a61b6bf4a7f8e6-Paper.pdf},
 volume = {32},
 year = {2019}
}

@inproceedings{chen2021spann,
 author = {Chen, Qi and Zhao, Bing and Wang, Haidong and Li, Mingqin and Liu, Chuanjie and Li, Zengzhong and Yang, Mao and Wang, Jingdong},
 booktitle = {Advances in Neural Information Processing Systems},
 editor = {M. Ranzato and A. Beygelzimer and Y. Dauphin and P.S. Liang and J. Wortman Vaughan},
 pages = {5199--5212},
 publisher = {Curran Associates, Inc.},
 title = {SPANN: Highly-efficient Billion-scale Approximate Nearest Neighborhood Search},
 url = {https://proceedings.neurips.cc/paper_files/paper/2021/file/299dc35e747eb77177d9cea10a802da2-Paper.pdf},
 volume = {34},
 year = {2021}
}

@INPROCEEDINGS{zhao2020song,
  author={Zhao, Weijie and Tan, Shulong and Li, Ping},
  booktitle={2020 IEEE 36th International Conference on Data Engineering (ICDE)}, 
  title={SONG: Approximate Nearest Neighbor Search on GPU}, 
  year={2020},
  volume={},
  number={},
  pages={1033-1044},
  doi={10.1109/ICDE48307.2020.00094}}

@INPROCEEDINGS{ootomo2024cagra,
  author={Ootomo, Hiroyuki and Naruse, Akira and Nolet, Corey and Wang, Ray and Feher, Tamas and Wang, Yong},
  booktitle={2024 IEEE 40th International Conference on Data Engineering (ICDE)}, 
  title={CAGRA: Highly Parallel Graph Construction and Approximate Nearest Neighbor Search for GPUs}, 
  year={2024},
  volume={},
  number={},
  pages={4236-4247},
  doi={10.1109/ICDE60146.2024.00323}}

@ARTICLE{khan2024bang,
  author={Venkatasubba, Karthik and Khan, Saim and Singh, Somesh and Simhadri, Harsha Vardhan and Vedurada, Jyothi},
  journal={IEEE Transactions on Big Data}, 
  title={BANG: Billion-Scale Approximate Nearest Neighbour Search Using a Single GPU}, 
  year={2025},
  volume={11},
  number={6},
  pages={3142-3157},
  doi={10.1109/TBDATA.2025.3581085}}

@inproceedings{liu2024juno,
author = {Liu, Zihan and Ni, Wentao and Leng, Jingwen and Feng, Yu and Guo, Cong and Chen, Quan and Li, Chao and Guo, Minyi and Zhu, Yuhao},
title = {JUNO: Optimizing High-Dimensional Approximate Nearest Neighbour Search with Sparsity-Aware Algorithm and Ray-Tracing Core Mapping},
year = {2024},
isbn = {9798400703850},
publisher = {Association for Computing Machinery},
address = {New York, NY, USA},
url = {https://doi.org/10.1145/3620665.3640360},
doi = {10.1145/3620665.3640360},
booktitle = {Proceedings of the 29th ACM International Conference on Architectural Support for Programming Languages and Operating Systems, Volume 2},
pages = {549–565},
numpages = {17},
location = {La Jolla, CA, USA},
series = {ASPLOS '24}
}

@INPROCEEDINGS{zhang2018fpga,
  author={Zhang, Jialiang and Li, Jing and Khoram, Soroosh},
  booktitle={2018 IEEE/CVF Conference on Computer Vision and Pattern Recognition}, 
  title={Efficient Large-Scale Approximate Nearest Neighbor Search on OpenCL FPGA}, 
  year={2018},
  volume={},
  number={},
  pages={4924-4932},
  doi={10.1109/CVPR.2018.00517}}

@INPROCEEDINGS{lee2022anna,
  author={Lee, Yejin and Choi, Hyunji and Min, Sunhong and Lee, Hyunseung and Beak, Sangwon and Jeong, Dawoon and Lee, Jae W. and Ham, Tae Jun},
  booktitle={2022 IEEE International Symposium on High-Performance Computer Architecture (HPCA)}, 
  title={ANNA: Specialized Architecture for Approximate Nearest Neighbor Search}, 
  year={2022},
  volume={},
  number={},
  pages={169-183},
  doi={10.1109/HPCA53966.2022.00021}}

@inproceedings{zeng2023dfgas,
author = {Zeng, Shulin and Zhu, Zhenhua and Liu, Jun and Zhang, Haoyu and Dai, Guohao and Zhou, Zixuan and Li, Shuangchen and Ning, Xuefei and Xie, Yuan and Yang, Huazhong and Wang, Yu},
title = {DF-GAS: a Distributed FPGA-as-a-Service Architecture towards Billion-Scale Graph-based Approximate Nearest Neighbor Search},
year = {2023},
isbn = {9798400703294},
publisher = {Association for Computing Machinery},
address = {New York, NY, USA},
url = {https://doi.org/10.1145/3613424.3614292},
doi = {10.1145/3613424.3614292},
booktitle = {Proceedings of the 56th Annual IEEE/ACM International Symposium on Microarchitecture},
pages = {283–296},
numpages = {14},
location = {Toronto, ON, Canada},
series = {MICRO '23}
}

@ARTICLE{yuan2025fanns,
  author={Yuan, Wei and Jin, Xi},
  journal={IEEE Transactions on Very Large Scale Integration (VLSI) Systems}, 
  title={FANNS: An FPGA-Based Approximate Nearest-Neighbor Search Accelerator}, 
  year={2025},
  volume={33},
  number={4},
  pages={1197-1201},
  doi={10.1109/TVLSI.2024.3496589}}

@misc{faiss,
      title={The Faiss library}, 
      author={Matthijs Douze and Alexandr Guzhva and Chengqi Deng and Jeff Johnson and Gergely Szilvasy and Pierre-Emmanuel Mazaré and Maria Lomeli and Lucas Hosseini and Hervé Jégou},
      year={2025},
      eprint={2401.08281},
      archivePrefix={arXiv},
      primaryClass={cs.LG},
      url={https://arxiv.org/abs/2401.08281}, 
}

@misc{faisscentered,
  author       = {{Faiss Developers}},
  title        = {{Faiss} Pull Request \#4550: Simplify {RaBitQ} Slightly, Improving Speed and Recall},
  howpublished = {GitHub pull request},
  year         = {2025},
  url          = {https://github.com/facebookresearch/faiss/pull/4550},
  note         = {Accessed: July 31, 2026}
}

@manual{upmem-doc,
  author       = {{UPMEM}},
  title        = {{UPMEM DPU SDK User Manual}},
  organization = {UPMEM},
  year         = {2025},
  version      = {2025.1.0},
  url          = {https://sdk.upmem.com/2025.1.0/},
  note         = {Accessed: July 31, 2026}
}

@manual{upmem2025codingtips,
  author       = {{UPMEM}},
  title        = {Coding Tips and Recommended Practices},
  organization = {UPMEM},
  year         = {2025},
  version      = {DPU SDK 2025.1.0},
  url          = {https://sdk.upmem.com/2025.1.0/fff_CodingTips.html},
  note         = {Accessed: July 31, 2026}
}

\end{document}